\documentclass[%
 reprint,showkeys,
superscriptaddress,
 amsmath,amssymb,
 aps,
pra,
floatfix,
]{revtex4-2}

\usepackage{graphicx}

\usepackage{dcolumn}
\usepackage{bm}
\usepackage{hyperref}

\newcommand{\ket}[1]{|#1\rangle}
\renewcommand{\hat}[1]{#1}

\DeclareMathOperator{\csch}{csch}

\begin{document}


\title{
Ancilla-Error-Transparent Controlled Beam Splitter Gate
}




\author{Iivari Pietik\"{a}inen}
\affiliation{Department of Optics, Palack\'{y} University, 17. listopadu 1192/12, 77146 Olomouc, Czechia}

\author{Ond\v{r}ej \v{C}ernot\'{i}k}
\affiliation{Department of Optics, Palack\'{y} University, 17. listopadu 1192/12, 77146 Olomouc, Czechia}

\author{Shruti Puri}

\affiliation{Yale Quantum Institute, PO Box 208 334, 17 Hillhouse Ave, New Haven, CT 06520-8263 USA }
\affiliation{Department of Applied Physics, Yale University}

\author{Radim Filip}
\affiliation{Department of Optics, Palack\'{y} University, 17. listopadu 1192/12, 77146 Olomouc, Czechia}

\author{S. M. Girvin}

\affiliation{Yale Quantum Institute, PO Box 208 334, 17 Hillhouse Ave, New Haven, CT 06520-8263 USA }
\affiliation{Department of Physics, Yale University}


\date{\today}

\begin{abstract}
In hybrid circuit QED architectures containing both ancilla qubits and bosonic modes, a controlled beam splitter gate is a powerful resource. It can be used to create (up to a {controlled-}parity operation) an ancilla-controlled SWAP gate acting on two bosonic modes. This 
is the essential element required to execute the `swap test' for purity, {prepare quantum non-Gaussian entanglement and directly measure nonlinear functionals} of quantum states. It also constitutes an important gate for hybrid discrete/continuous-variable quantum computation. We propose a new realization of a hybrid cSWAP utilizing `Kerr-cat' qubits\textemdash anharmonic oscillators subject to strong two-photon driving. The Kerr-cat is used to generate a controlled-phase beam splitter (cPBS) operation. When combined with an ordinary beam splitter one obtains a controlled beam-splitter (cBS) and from this a cSWAP. The strongly biased error channel for the Kerr-cat has phase flips which dominate over bit flips. This yields important benefits for the cSWAP gate which becomes non-destructive and transparent to the dominate error. {Our proposal is straightforward to implement and, based on currently existing experimental parameters, should achieve controlled beam-splitter gates with high fidelities comparable to current ordinary beam-splitter operations available in circuit QED.}
\end{abstract}

\maketitle


\section{Introduction}






In this work we propose a new scheme for realizing a hybrid discrete/continuous-variable controlled-SWAP (or Fredkin) gate which, conditioned on the quantum state of an ancilla qubit C,  applies a beam splitter operation that can be used to swap the quantum states of two bosonic modes A and B ($a\rightarrow -b, b\rightarrow +a$).
A key advantage of our approach is that it utilizes a Kerr-cat qubit \cite{Puri2017,Puri2019,Puri2020} both as a noise-biased control ancilla and as the driven non-linear element that creates the linear beam-splitter. This innovation renders the gate error-transparent to the dominant error channel (ancilla dephasing associated with excitation loss) { so that ancilla errors do not propagate into the data modes. In addition, we predict that this new protocol yields a gate time that can be substantially shorter than the existing circuit QED protocol for cSWAP \cite{Gao2019}.}

In the most general context, the pair of systems \{A, B\} could be either discrete variable (qubits) or continuous variable (bosonic modes). In the discrete variable context, the cSWAP is a non-Clifford gate (a member of the second level of the Clifford hierarchy) with important applications for universal quantum computation, including machine learning \cite{WeedbrookQML2017,PhysRevResearch.1.033159,PhysRevA.101.052309}, for routing quantum information through a quantum-controlled switching network to create a quantum random access memory (QRAM) \cite{LloydQRAM2008,PhysRevA.78.052310,PhysRevA.86.010306,arunachalam2015,%
8962352,PhysRevA.102.032608,QRAMConnor2021}, and, in general, for state preparation of quantum non-Gaussian entanglement \cite{Gerry_PhysRevA.59.4095,PhysRevA.65.043802} and carrying out the `swap test' for the purity of a quantum state \cite{PhysRevA.65.062320,nguyen2021experimentalSWAP}, computing the Renyi entropy \cite{RenyiEntropy_PhysRevA.98.052334} or the overlap of two different quantum states for quantum fingerprinting \cite{PhysRevLett.87.167902} and other verification purposes \cite{PhysRevA.65.062320,PRXQuantum.2.010102},  and a variety of related tasks \cite{VisualTrackingPhysRevA.99.022301,QADC_PhysRevA.99.012301,QEM_PhysRevX.8.031027,%
Suba__2019,PhysRevLett.88.217901,PhysRevLett.125.120502,PhysRevLett.124.010506}.  

Non-deterministic cSWAP gates have been achieved in photonic systems \cite{Patel2016,Ono2017,Starek2018_non-destructive} and deterministic cSWAP of bosonic modes controlled by a qubit has been achieved in superconducting circuits \cite{Gao2019} and in ion traps \cite{Modularquantumcomputationtrappedionsystem,Gan_PhysRevLett.124.170502}.  Deterministic cSWAP is  a key element in circuits used in the experimental realization of exponential SWAP (eSWAP) gates \cite{Gao2019}. Lau and Plenio \cite{Lau_Plenio_PhysRevLett.117.100501} have shown that eSWAP can be used for universal computation using bosonic modes. The Lau and Plenio scheme offers the important feature that different (error correctable) bosonic encodings can be used without changing the universal instruction set architecture, since SWAP and eSWAP are agnostic to the contents of the bosonic modes being swapped.

For qubits, the gate set \{cSWAP,CNOT,Hadamard\} is equivalent to \{Toffoli,Hadamard\} which is universal \cite{Aharanov2003}. 
For both discrete and continuous variables, cSWAP finds powerful application in modularizing quantum computation \cite{Modularquantumcomputationtrappedionsystem} and can be used to turn an arbitrary unknown unitary into a controlled unitary  \cite{Obrien_Modular}. 

We focus here on the hybrid circuit QED architecture \cite{Blais2020,BlaiscQEDReviewRMP2020,Krantz2019} which contains both discrete-variable (DV) components (e.g., transmon \cite{Koch2007,Schuster2007a,Lifeafterchargenoise,Paik3D_PhysRevLett.107.240501} or Kerr-cat qubits \cite{Puri2017,Puri2019,Puri2020}) and continuous-variable (CV) components containing bosonic modes (e.g., microwave \cite{Gao2019,Wang2020FCFs,Campagne2020,Reinhold2020ErrorCorrectedGate} or mechanical resonators \cite{Lhaye2009,OConnell2010,Gustafsson2012,Pirkkalainen2013,Gustafsson2014,Rouxinol2016,Manenti2017,Yiwen_Chu2017,YiwenPhononFock2018,Delsing2019_PhysRevA.99.013840,Arrangoiz-Arriola2019,Mirhosseini2020}). In such a hybrid architecture one can have gates such as cSWAP acting purely within the DV sector or acting on the CV sector but controlled by the DV sector.  An open challenge in the field is to develop cSWAP acting entirely within the CV sector.

\begin{figure}[b]
\includegraphics[width=0.4\textwidth]{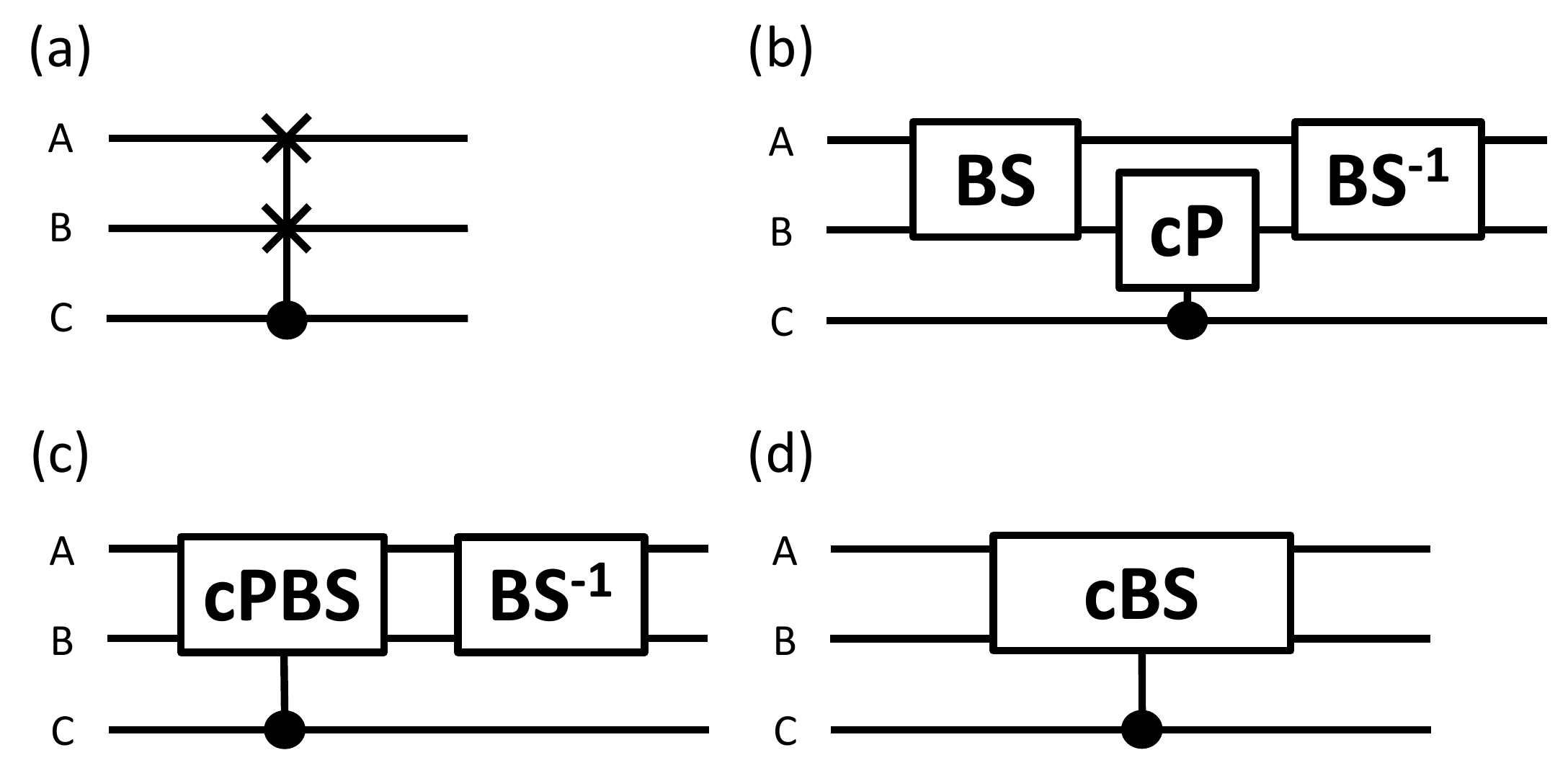}
\caption{\label{fig:cSWAPcircuit} (a) Standard circuit symbol for the controlled SWAP (cSWAP) gate.  (b) Mach-Zehnder interferometer circuit for realization of cSWAP between two bosonic modes. BS and BS$^{-1}$ are 50:50 beam splitter unitaries. The ancilla is a transmon qubit that applies a controlled parity unitary (cP), shifting the phase in one arm of the interferometer by $\pi$ if the ancilla is in $|1\rangle$.  (c) The new cSWAP circuit proposed in this work. Here the ancilla is a Kerr-cat qubit. The first gate, cPBS, is a controlled-phase beam splitter which produces BS if the ancilla is in $|0\rangle$ and BS$^{-1}$ if the ancilla is in $|1\rangle$.   (d) Representation of the gate shown in (c) as a single controlled-SWAP beam splitter (cBS) unitary which applies the identity if the ancilla is in $|0\rangle$ and applies a SWAP if the ancilla is in  $|1\rangle$. Changing the duration of the beam splitter and cPBS gates allows creation of general controlled beam splitter operations of which cSWAP is a special case {(up to a controlled-parity on one of the modes)}.
\label{fig:cSWAP_figure}}
\end{figure}

In circuit QED, a deterministic cSWAP gate between two microwave resonator modes was achieved by Gao et al. \cite{Gao2019} using a scheme based on the differential dispersive shift of two cavities coupled to the same transmon qubit.  Another scheme that could in principle be used is the `temporal Mach-Zehnder interferometer' circuit shown in Fig.~\ref{fig:cSWAP_figure}b.  In this scheme, the modes being swapped are bosonic modes stored in microwave cavities but the control mode C is a transmon qubit.  The control mode is used to apply a controlled-parity gate (cPHASE gate with phase $\pi$) which does nothing if C is in state $|0\rangle$ but shifts the phase difference between the two arms of the interferometer by $\pi$ if C is in state $|1\rangle$.  The interference between the two paths of the interferometer then results in IDENTITY or SWAP, conditioned on the state of C.  An extension of this interferometeric scheme was used by Gao et al.\ \cite{Gao2019} to achieve an exponential SWAP (eSWAP) gate.

In this work we propose a new interferometric scheme for cSWAP in which the controlled phase is applied, not to one of the bosonic modes in the interferometer, but rather to one of the beam splitters as shown in Fig.~\ref{fig:cSWAP_figure}c.  When such a controlled-phase beam splitter (cPBS) is combined with an ordinary beam splitter (BS), the result is a controlled beam splitter (cBS) Hamiltonian, turned on and off by the state of C.  Appropriately choosing the duration of this gate  yields cSWAP.  The cBS Hamiltonian would also be useful in realizing simulation of non-trivial quantum Hamiltonian models involving spins (or fermions) coupled to bosons. 

We propose to realize a controlled-phase beam splitter through use of the Kerr-cat qubit, 
a DV component that, unlike previous transmon implementations, features a highly-biased noise channel (in which bit flips {require overcoming a large barrier and thus} are much rarer than phase flips { which are associated with energy damping}) \cite{Puri2017,Puri2019,Puri2020,Grimm2020}.  Furthermore, unlike ordinary DV qubits, the Kerr-cat has an underlying continuous {rotation} symmetry which allows one to escape a no-go theorem that otherwise prevents creation of a cNOT gate that preserves the noise bias~\cite{Puri2020}.  These two features significantly improve error-correction thresholds for circuits constructed from Kerr-cats \cite{Grimm2020,darmawan2021practical}.

It was shown in Ref.~\cite{Puri2019} that the Kerr-cat qubit could be used as a fault-tolerant error syndrome detector for a variety of codes. For the GKP bosonic codes this is realized in the form of ancilla-controlled oscillator displacements \cite{Campagne2020}
\begin{equation}
    D_\mathrm{c}(\beta)=e^{Z[\beta a^\dagger -\beta^* a]},
\end{equation}
where $\beta$ is a complex number representing  the dimensionless displacement in phase space {and $Z$ is the Pauli operator of the Kerr-cat qubit}.
As we will demonstrate, an appropriately driven Kerr-cat qubit can also yield an effective beam-splitter Hamiltonian between two bosonic modes $a$ and $b$ of the form 
\begin{equation}\label{eq:HBSIntro}
    H_\mathrm{BS}=i\alpha Z[\lambda(t) a^\dagger b-\lambda^*(t)^* ab^\dagger],
\end{equation}
where $\lambda(t)$ is the complex envelope amplitude of a special pump tone, and {$\pm\alpha$} is the (assumed real) amplitude of the spontaneous coherent state {$|\pm\alpha\rangle$} formed by the Kerr cat.  This has opposite sign in the two standard basis states (eigenstates of $Z$) of the qubit, and thereby controls the phase of the beam splitter.  We will show that the physical origin of the controlled phase beam splitter Hamiltonian \eqref{eq:HBSIntro} is the four-wave mixing among the two cavity modes, the spontaneous oscillation of the Kerr-cat qubit and the pump tone.

The unitary evolution operator of the system over the interval in which the beam splitter \eqref{eq:HBSIntro} is turned on is given (for the case that $\lambda(t)$ is real) by
\begin{equation}
    U_\mathrm{c}(\theta)=e^{\frac{\theta}{2}Z[a^\dagger b-ab^\dagger]},
\end{equation}
where 
\begin{equation}
    \theta = \alpha\int dt\, \lambda(t),
\end{equation}
{and throughout the paper, unless otherwise stated, we are working in a rotating frame (interaction picture) in which both of the bare oscillator frequencies (which are different from each other in the lab frame) are zero.}
For $\theta=\pi/2$ we have a (conditional phase) 50:50 beam splitter and for $\theta=\pi$ we have (up to a phase) a SWAP gate that sends
\begin{eqnarray}
a&\rightarrow& -b\\
b&\rightarrow&+a.
\end{eqnarray}
By combining this { 50:50 cPBS with an ordinary balanced (50:50) beam splitter} 
\begin{equation}
    U(\theta)=e^{\frac{\theta}{2}[a^\dagger b-ab^\dagger]},
\end{equation}
we obtain a cSWAP gate
\begin{equation}
    \mathrm{cSWAP}=U(\frac{\pi}{2})U_\mathrm{c}(-\frac{\pi}{2})
    =e^{\frac{\pi}{4}[\hat I - Z][a^\dagger b-ab^\dagger]},
\end{equation}
which yields the identity for $Z=+1$ and SWAP (up to the phase mentioned above) for $Z=-1$.

Notice that because a bit flip reverses the phase of the beam splitter Hamiltonian, a bit flip error during the gate reduces $\theta$ below its intended value.  Thus if performing a cSWAP on (say) two error-correctable bosonic code words, an ancilla bit flip will lead to logical errors and (possibly) leakage errors out of the bosonic code space.  As a simple example of the latter, consider an (uncorrectable) bosonic encoding that represents logical 0 by 0 bosons and logical 1 by one boson.  Ideally
\begin{equation}
    \mathrm{cSWAP}[c_0\ket{0}+c_1\ket{1}]|11\rangle=[c_0\ket{0}-c_1\ket{1}]|11\rangle,
\end{equation}
where the term in square brackets denotes the state of the ancilla. However, an ancilla {bit-flip} error {caused by dissipation} will change the beam splitter ratio, leading, via the Hong-Ou Mandel effect \cite{GaoYY2018}, to leakage out of the code space
\begin{equation}
    |11\rangle\rightarrow \eta |11\rangle + \mu\frac{1}{\sqrt{2}}[|02\rangle+|20\rangle],
\end{equation}
where the coefficients $\eta,\mu$ depend on the precise time within the gate duration at which the ancilla error occurs.

Conversely, $U_\mathrm{c}(\theta)$ is error-transparent with respect to phase flips since $[U_\mathrm{c}(\theta),Z]=0.$  These facts suggest that the use of a highly noise-biased ancilla could be beneficial.  The fact that the Kerr-cat qubit exponentially suppresses bit flips at the cost of a only modest linear increase in phase flips
\cite{Puri2017,Puri2019,Puri2020}
is therefore an important feature. 

While this paper is focused on using the Kerr-Cat qubit in a hybrid CV-DV architecture as an ancilla for controlling the quantum states of microwave resonators, the ideas presented here can also be applied to {a DV} architecture solely based on Kerr-cat qubits. Ref.~\cite{Puri2020} shows how to create a bias-preserving $ZZ(\theta)$ gate between Kerr cats.  The ideas we present here would permit a bias-preserving $ZZZ(\theta)$ gate among three Kerr cat qubits, each having different frequencies.

    The error-transparency of the present gate construction to the dominant error source of the Kerr-cat ancilla enables the realization of an (almost) nondestructive measurement of the SWAP operator or equivalently, the exchange symmetry between multi-qubit or photonic systems. This is because the dominant phase-flip error in the ancilla which applies the conditional-SWAP, can only cause a misidentification of the exchange symmetry of the states, and cannot change or destroy the states being swapped {\cite{Cernotik2021}}. This is unlike the case of a controlled-SWAP operation with an ancilla suffering from bit-flip errors since this type of error results in destructive  back action on the bosonic states. This is because a bit-flip error in the middle of the SWAP leads to an incomplete swap of the two modes.  A nondestructive measurement of the SWAP operator considerably simplifies protocols for stabilization of quantum computations~\cite{berthiaume1994stabilisation,barenco1997stabilization,peres1999error}, state purification~\cite{cirac1999optimal} and cooling~\cite{cotler2019quantum}. Note that, a measurement of the SWAP operator is also useful for SWAP tests to measure the distinguishability of two input states \cite{Gao2019}. However, in this case a destructive SWAP measurement suffices as fresh input states are fed into the protocol. A non-destructive SWAP measurement is highly desirable when the post-measurement states are required for subsequent operations in the algorithm and cannot be simply discarded.

\section{Controlled-beam splitter with a Kerr-cat ancilla}

\subsection{The cBS Hamiltonian}

The full Hamiltonian of the system (in the lab frame) consisting of a driven SNAIL device coupled to two cavity fields is
\begin{subequations}\label{eq:Hini}
\begin{equation}
\hat{H} = \hat{H}_{\rm fields} +\hat{H}_{\rm SNAIL} +\hat{H}_{\rm drive},
\end{equation}
\begin{eqnarray}
\hat{H}_{\rm fields} &=& \omega_{a} \hat{a}^\dagger\hat{a} +\omega_{b} \hat{b}^\dagger\hat{b} +g_a(\hat{a}^\dagger\hat{c} +\hat{c}^\dagger\hat{a}) \nonumber\\ &&+g_b(\hat{b}^\dagger\hat{c} +\hat{c}^\dagger\hat{b}), \\ 
\hat{H}_{\rm SNAIL} &=& \omega_{c} \hat{c}^\dagger\hat{c} +g_3(\hat{c}^\dagger +\hat{c})^3 +g_4(\hat{c}^\dagger +\hat{c})^4, \\ 
\hat{H}_{\rm drive} &=& \sum_k (e^{-i\omega_k t}\epsilon_k\hat{c}^\dagger +e^{i\omega_k t}\epsilon_k^*\hat{c}).
\end{eqnarray}
\end{subequations}
The Hamiltonian $\hat{H}_{\rm fields}$ describes the evolution of the fields (annihilation operators $\hat{a}$, $\hat{b}$ and frequencies $\omega_{a,b}$) coupled to the SNAIL (annihilation operator $\hat{c}$) at rates $g_{a,b}$.
Next, the Hamiltonian $\hat{H}_{\rm SNAIL}$ describes the SNAIL with frequency $\omega_{c}$ and cubic and quartic nonlinearities $g_{3,4}$. Finally, the Hamiltonian $\hat{H}_{\rm drive}$ describes driving of the SNAIL with tones at frequencies $\omega_k$ and with amplitudes $\epsilon_k$.

\begin{figure}
    \centering
    \includegraphics[width=\linewidth]{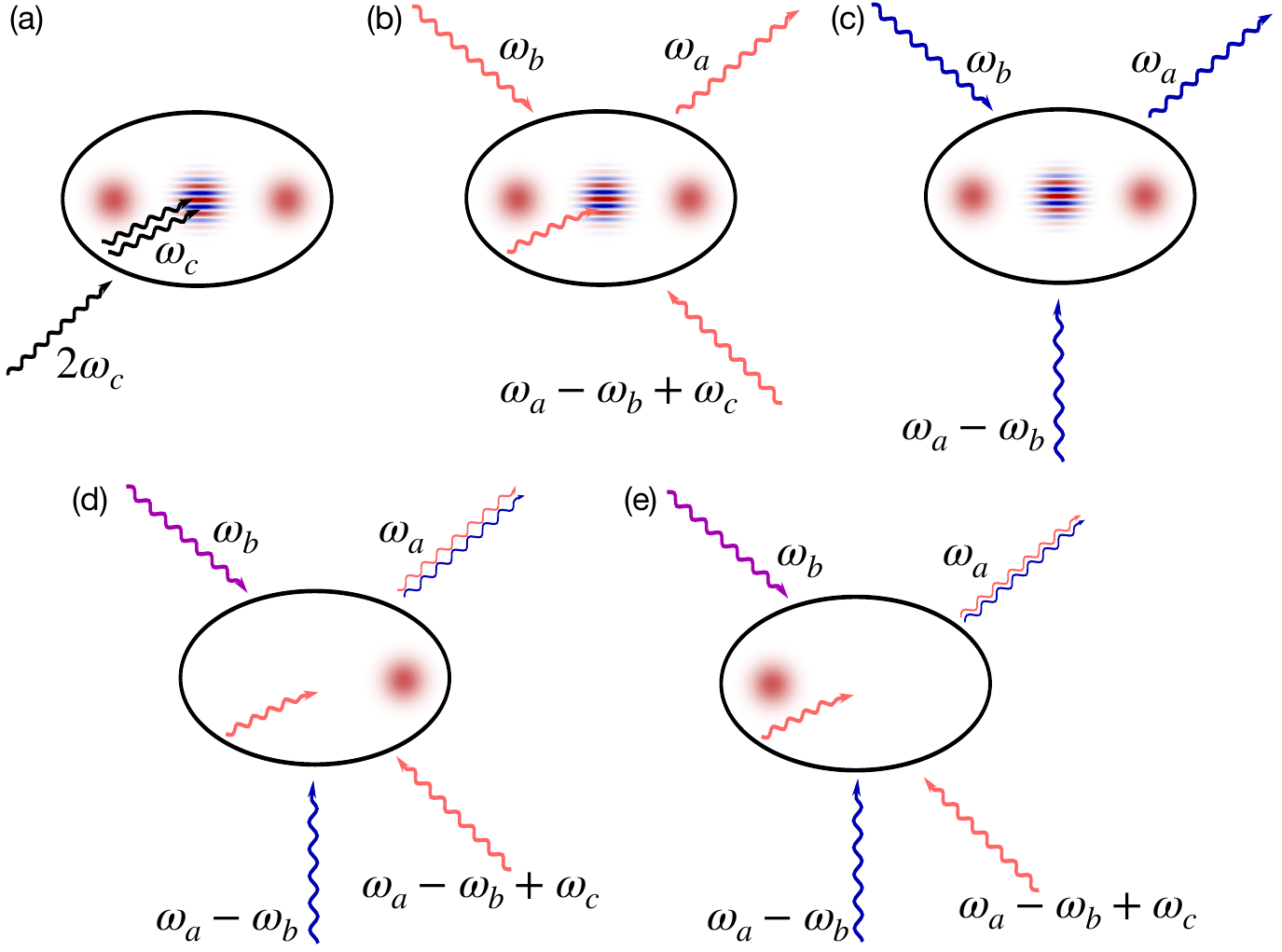}
    \caption{\label{fig:mixing}(a) Three-wave mixing process to create the two-photon driving that realizes the Kerr-cat logical space, { $\mathrm{Span}\{|\mathcal{C}_+\rangle,|\mathcal{C}_-\rangle\}$}.  Wigner function of the even-parity cat state is shown. (b) Controlled-phase beam splitter (cPBS) created via four-wave mixing using a single external pump plus a pump supplied by  the internal oscillation state of the Kerr-cat. (c) Beam-splitter (BS) coupling between frequencies $\omega_a$ and $\omega_b$ created via driving the three-wave mixing at frequency $\Delta=\omega_a-\omega_b$. Note that this process is independent of the internal state of the Kerr-cat. (d) For the cat qubit in the logical state $\ket{0}$ (the coherent state $\ket{\alpha}$), the 50:50 cPBS and BS interactions interfere destructively and no swapping takes place. (e) 50:50 BS and 50:50 cPBS with the cat qubit in the logical state $\ket{1}$ (the coherent state $\ket{-\alpha}$) combine to perform a swap between the cavity fields $a$ and $b$.}
\end{figure}

\begin{figure}[b]
\includegraphics[width=0.4\textwidth]{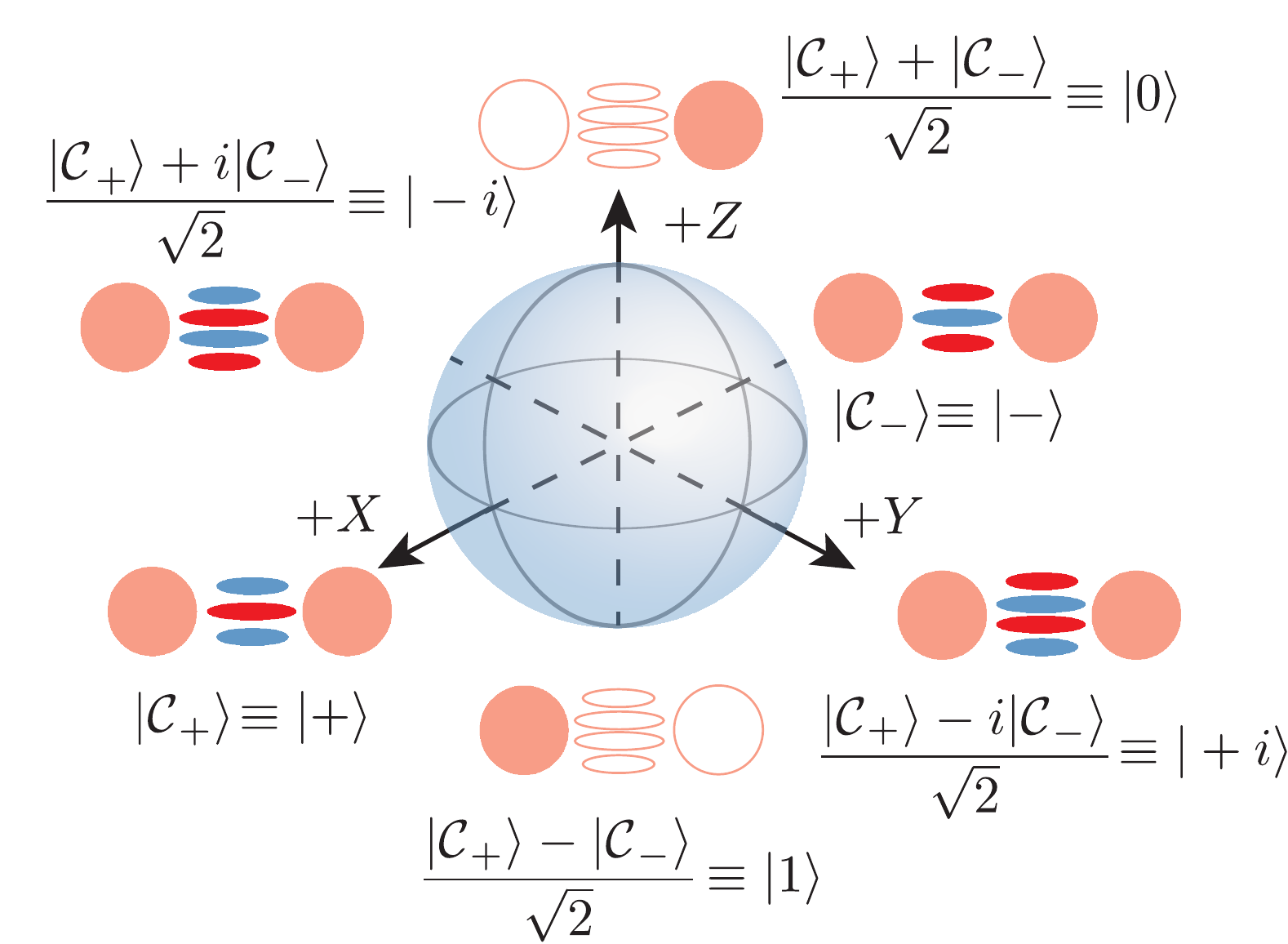}
\caption{\label{fig:bloch} Illustration of the Bloch sphere. The cat states $\ket{\mathcal{C}_\pm}=\mathcal{N}_\pm(\ket{\alpha}\pm\ket{-\alpha})$ are aligned along the $X$-axis of the Bloch sphere. Here, $\mathcal{N}_\pm$ are the normalization constants. The superpositions $(\ket{\mathcal{C}_+}\pm\ket{\mathcal{C}_-})/\sqrt{2}\simeq\ket{\pm\alpha}$ are aligned along the $Z$-axis (where the approximations hold for large $|\alpha|$).  The $|\pm \rangle$ eigenstates have definite photon number parity and thus single photon loss (or gain) acts (primarily) as a Pauli $Z$ error that causes dephasing which flips $|+\rangle$ to $|-\rangle$ and vice versa (and similarly for $|\pm i\rangle$). }
\end{figure}

The third- and fourth-order nonlinearities in the Hamiltonian $\hat{H}_{\rm SNAIL}$ create three- and four-wave mixing processes in the system~\cite{Frattini2017,Grimm2020}. With suitable driving frequencies, these processes can be used to generate specific interactions between the fields and the SNAIL with processes relevant for the {controlled beam splitter} gate shown schematically in Fig.~\ref{fig:mixing}. First, driving the three-wave mixer with the coupling strength $g_3$ at frequency $\omega_1 = 2\omega_c$ [panel (a)] leads to two-photon driving of the SNAIL which, in combination with the Kerr nonlinearity {(stemming from the quartic nonlinearity of the SNAIL)}, creates and stabilizes the Kerr-cat qubit~\cite{Grimm2020},
\begin{equation}
    \hat{H}_{\rm cat} = -K\hat{c}^{\dagger 2}\hat{c}^2 + \epsilon\hat{c}^{\dagger 2} + \epsilon^\ast\hat{c}^2.
\end{equation}
The Hamiltonian $\hat{H}_{\rm cat}$ has two degenerate ground states, the cat states $\ket{\mathcal{C}_\pm}=\mathcal{N}_\pm(\ket{\alpha}\pm\ket{-\alpha})$, with $\alpha = \sqrt{\epsilon/K}$, where $\ket{\pm\alpha}$ are coherent states, and $\mathcal{N}_\pm$ are the normalization constants.  The two cat states are orthogonal and can be used to encode a qubit. The Bloch sphere used here is shown in Fig.~\ref{fig:bloch}. The computational states $\ket{0},\ket{1}$ are taken to be superpositions of the two cat states $(\ket{\mathcal{C}_+}\pm\ket{\mathcal{C}_-})/\sqrt{2}\simeq\ket{\pm\alpha}$, where the approximations holds for large $|\alpha|$. 
The strength of the Kerr nonlinearity $K$ and of the two-photon drive $\epsilon$ are related to the strength of the pump and the nonlinearity $g_4$ as described in Appendix~\ref{app:Hamiltonian}.

To create a beam splitter between the cavity fields $\hat{a}$ and $\hat{b}$ with a phase that depends on the state of the Kerr-cat qubit, we employ a four-wave mixing process (coupling rate $g_4$) where one of the two pumps is provided by the spontaneous oscillation of the Kerr cat itself at frequency $\omega_c$. 
{We set the frequency of the second pump to be $\omega_2 = \omega_a-\omega_b+\omega_c$ [see also Fig.~\ref{fig:mixing}(b)] which, as described in Appendix~\ref{app:Hamiltonian}, gives rise to the cPBS Hamiltonian (again in the rotating frame)}
\begin{equation}\label{eq:cPBS}
    \hat{H}_{\rm cPBS} = -\zeta_1\hat{a}^\dagger\hat{b}\hat{c}^\dagger - \zeta_1^\ast\hat{a}\hat{b}^\dagger\hat{c}
\end{equation}
at a rate $\zeta_1$.
To see how this Hamiltonian gives rise to a beam splitter with a phase controlled by the Kerr cat, we consider the Kerr cat in a mean-field approximation, $\langle\hat{c}\rangle=\langle\hat{c}^\dagger\rangle=\pm\alpha$.
In this approximation (valid for moderately large amplitudes, $\alpha\gtrsim\sqrt{3}$), the cPBS Hamiltonian becomes
\begin{equation}\label{eq:MF}
\hat{H}_{\rm mf} = \mp\zeta_1\alpha\hat{a}^\dagger\hat{b} \mp\zeta_1^\ast\alpha\hat{a}\hat{b}^\dagger,
\end{equation}
with the $-$ ($+$) sign corresponding to the Kerr-cat qubit state $|0\rangle$ ($|1\rangle$).
This Hamiltonian transforms the cavity fields according to
\begin{subequations}\label{eq:CPBSideal}
\begin{align}
	a(t) &= \cos(\zeta_0 t)a_0 + i\sin(\zeta_0 t)e^{i\phi_\pm}b_0, \\
	b(t) &= i\sin(\zeta_0 t)e^{-i\phi_\pm}a_0 + \cos(\zeta_0 t)b_0,
\end{align}
\end{subequations}
where we separated the beam-splitter rate into its amplitude and phase $\mp\zeta_1\alpha = \zeta_0 e^{i\phi_\pm}$ and we denote the initial states of the fields by $a_0,b_0$.

Since the phases acquired by the fields during the cPBS interaction differ by $\pi$, $\phi_+ = \phi_-+\pi$, the two processes (corresponding to the { Kerr-cat qubit in one of the logical states $\ket{0},\ket{1}$}) are Hermitian conjugates (and thus inverses) of each other. We can therefore turn the controlled-phase beam splitter into a controlled beam splitter by using an additional beam-splitter interaction which can be engineered by pumping a three-wave mixing process at frequency $\Delta =\omega_a-\omega_b$ [Fig.~\ref{fig:mixing}(c)] or, alternatively, using four-wave mixing with two drive tones at frequencies $\omega_{3,4}$ satisfying $\omega_3-\omega_4 = \omega_a-\omega_b$~\cite{Gao2019,Zhang2019} {(see also appendix \ref{app:schemes} for discussion of alternative driving schemes)}.
We thus obtain the beam-splitter Hamiltonian
\begin{equation}
    \hat{H}_{\rm BS} = \zeta_2\hat{a}^\dagger\hat{b}+\zeta_2^\ast\hat{a}\hat{b}^\dagger,
\end{equation}
where the phase of the interaction constant $\zeta_2$ can be controlled by the phase of the pump field.
When we set the phases of the interactions and gate times $t_{1,2}$ such that $\zeta_1\alpha t_1 = \zeta_2 t_2$ (where $t_{1,2}$ is the total time of the cPBS and BS interaction, respectively), the beam splitter and controlled-phase beam splitter cancel each other for the { qubit state $\ket{0}$ [see Fig.~\ref{fig:mixing}(d)].
On the other hand, for the cat in the state $\ket{1}$,} the fields are transformed as [see also Fig.~\ref{fig:mixing}(e)]
\begin{subequations}
\begin{align}
	a(t) &= \cos(2\zeta_1\alpha t_1)a_0 -i\sin(2\zeta_1\alpha t_1)e^{i\phi_+}b_0, \\
	b(t) &= -i\sin(2\zeta_1\alpha t_1)e^{-i\phi_+}a_0 + \cos(2\zeta_1\alpha t_1)b_0.
\end{align}
\end{subequations}
These transformations give the controlled-beam splitter gate. We can engineer any {(controlled)} splitting ratio between the cavity fields by fixing the drive times for the cPBS and BS couplings; 
for a full swap between the cavities, both cPBS and BS are 50:50 {beam splitters (with either the same or opposite phases)}. 


The final interaction, {which is always present and limits} the gate fidelity, is the cross-Kerr interaction between the SNAIL and the cavity fields (see Appendix~\ref{app:Hamiltonian}),
\begin{equation}
    \hat{H}_{\rm cK} = -(\chi_a\hat{a}^\dagger\hat{a}+\chi_b\hat{b}^\dagger\hat{b})\hat{c}^\dagger\hat{c},
\end{equation}
which introduces a frequency shift on the cavity fields proportional to the number of photons in the SNAIL or, equivalently, a frequency shift on the SNAIL proportional to the total number of photons in the two cavity fields.
To minimize its effect, we compensate the mean-field part of the cross-Kerr interaction by suitably shifting the frequency of the rotating frame and all drives as described in Appendix~\ref{app:Kerr}.
The cross-Kerr Hamiltonian then becomes
\begin{equation}
    \hat{H}_{\rm cK} = -\chi(\hat{a}^\dagger\hat{a}+\hat{b}^\dagger\hat{b}-N)(\hat{c}^\dagger\hat{c}-|\alpha|^2),
\end{equation}
where $N = \langle\hat{a}^\dagger\hat{a}+\hat{b}^\dagger\hat{b}\rangle$ is the mean photon number of the two cavity fields and $|\alpha|^2$ is the average occupation of the cat;
we also assumed that the two cavity fields have the same cross-Kerr interaction, $\chi_a=\chi_b=\chi$.

Together, all these interactions give rise to the total effective Hamiltonian
\begin{equation}\label{eq:HamEff}
\begin{split}
\hat{H} &= \hat{H}_{\rm cat} + \hat{H}_{\rm cPBS} + \hat{H}_{\rm BS} + \hat{H}_{\rm cK} \\
&= -K\hat{c}^{\dagger 2}\hat{c}^2 +\epsilon\hat{c}^{\dagger 2} +\epsilon^* \hat{c}^{2}\\
&\quad -\zeta_1\hat{a}^\dagger\hat{b}\hat{c}^\dagger -\zeta_1^\ast\hat{a}\hat{b}^\dagger\hat{c} +\zeta_2\hat{a}^\dagger\hat{b} + \zeta_2^\ast \hat{a}\hat{b}^\dagger\\
&\quad -\chi (\hat{a}^\dagger\hat{a} + \hat{b}^\dagger\hat{b}-N)(\hat{c}^\dagger\hat{c}-|\alpha|^2).
\end{split}
\end{equation}
In the cPBS and BS interactions, the coefficients $\zeta_{1,2}$ are now time-dependent to account for switching the interactions on and off in accordance with the interferometric scheme in Fig.~\ref{fig:cSWAP_figure}(c).

\subsection{Noise Bias in controlled beam splitter}

It has been shown that if noise causes only small displacements of states in phase space then the noise-channel of the Kerr cat is biased so that bit-flip errors are strongly suppressed compared to phase-flip errors \cite{Puri2017,Puri2019,Puri2020}. Under such a reasonable assumption about practical environmental noise, the probability of a non-dephasing or bit-flip type error decreases exponentially with the size of the cat $|\alpha|^2$, while the probability of a phase-flip error increases polynomially with the cat size. When the dominant source of noise is single-photon loss, the rate of phase-flip error scales as $O(|\alpha|^2)$ while the cBS gate time (for fixed $\zeta$) scales as $O(1/|\alpha|)$ Thus, the probability of a phase-flip error increases only linearly with $|\alpha|$. In principle, the {external pump amplitude (required to activate the cPBS)} and hence $\zeta_{1,2}$ is limited by the energy gap of the Kerr-cat qubit which itself increases with  $|\alpha|^2$. Thus, it may even be possible to reduce the phase-flip error probability by going to larger amplitude cat. In summary, the strong bias available along with {the relatively} low probability of phase-flip errors makes the Kerr-cat a promising candidate for mediating a controlled beam-splitter operation between two oscillator modes. 






Two-photon dissipation can be added to help stabilize the Kerr-cat {against leakage errors}.  This dissipation cools the Kerr-cat back into the logical manifold if a leakage error occurs \cite{Puri2020} and cat states can also be stabilized purely by two-photon dissipation without using the Kerr effect \cite{TouzardDissCat_PhysRevX.8.021005,LeghtasDissipativeCat2020}. Two-photon dissipation  commutes with the photon number parity operator and hence does not cause dephasing errors.  Recently a `colored cat qubit' \cite{putterman2021colored} that is stabilized by single-photon dissipation has been proposed, but we have not included this possibility in our simulations.

The dissipative dynamics discussed above are modeled by the master equation
\begin{equation}\label{eq:ME}
\dot{\hat{\rho}} = -i[\hat{H},\hat{\rho}] +\kappa(1+N_t)\mathcal{D}[\hat{c}]\hat{\rho} +\kappa N_t\mathcal{D}[\hat{c}^\dagger]\hat{\rho} +\kappa_2\mathcal{D}[\hat{c}^2]\hat{\rho} \,,
\end{equation}
 with Hamiltonian (\ref{eq:HamEff}). Here $\mathcal{D}[\hat{o}]\hat{\rho} = \hat{o}\hat{\rho}\hat{o}^\dagger -\frac{1}{2}\hat{o}^\dagger\hat{o}\hat{\rho} -\frac{1}{2}\hat{\rho}\hat{o}^\dagger\hat{o}$ is the Lindblad superoperator, $N_t$ is the thermal population of the Kerr cat mode, and $\kappa$, $\kappa_2$ are the single- and two-photon dissipation rates of the ancilla. The single-photon loss and gain of the SNAIL mode (the first two Lindblad superoperators) stem from interactions with the intrinsic reservoir and explain well the experimental observations \cite{Grimm2020}. We include in addition two-photon dissipation (the last Lindblad superoperator) to help stabilize the Kerr-cat qubit as described above. 
 For all our calculations we find that infidelities are dominated by ancilla errors {(caused by damping and dephasing)} and we neglect the intrinsic damping of the bosonic modes which is small in comparison. Because of the coupling of the bosonic modes to the ancilla, they suffer an additional `inverse Purcell' damping~\cite{Reagor2016} $(g_i/\Delta_i)^2\kappa = 0.0225\kappa$ (see Table~\ref{table}) which we also neglect.
 
 
\begin{table}
\begin{tabular}{|l|l|l|l|l|}
\hline
&Refs.\ \cite{GaoYY2018,Gao2019} &Ref.\ \cite{Grimm2020}  &Ref.\ \cite{Eickbusch2021X} &This work \\
\hline
$\rm{Kerr}/(2\pi)$ &71.25 MHz &6.7 MHz &96.5 MHz &6.7 MHz \\
$\chi_{a,b}/(2\pi)$ & 370, 300 kHz &200-250 kHz &33 kHz &600 kHz \\
$g_{a,b}/\Delta_{a,b}$ &0.036, 0.032 &0.086-0.097 &0.0092 &0.15 \\
$\xi_{1,2}$ &$\sim$ 0.2, 0.4 &0.15-0.16 & &0.2 \\
$\alpha$ & &$\sim\sqrt{3}$ & &$\sqrt{3}$ \\
$t_{\rm SWAP}$ &$\sim 10 \;{\rm \mu s}$ & & &$1.2 \;{\rm \mu s}$ \\
\hline
\end{tabular}
\caption{Comparison of the system parameters used in our simulations with the experiments of Gao et al. \cite{Gao2019}, Grimm et al. \cite{Grimm2020} and Eickbusch et al. \cite{Eickbusch2021X}. Empty rows indicate that the given parameter has not been used in the corresponding experiment. }\label{table}
\end{table}

\begin{figure}
\begin{center}
	\includegraphics[width=1.0\linewidth]{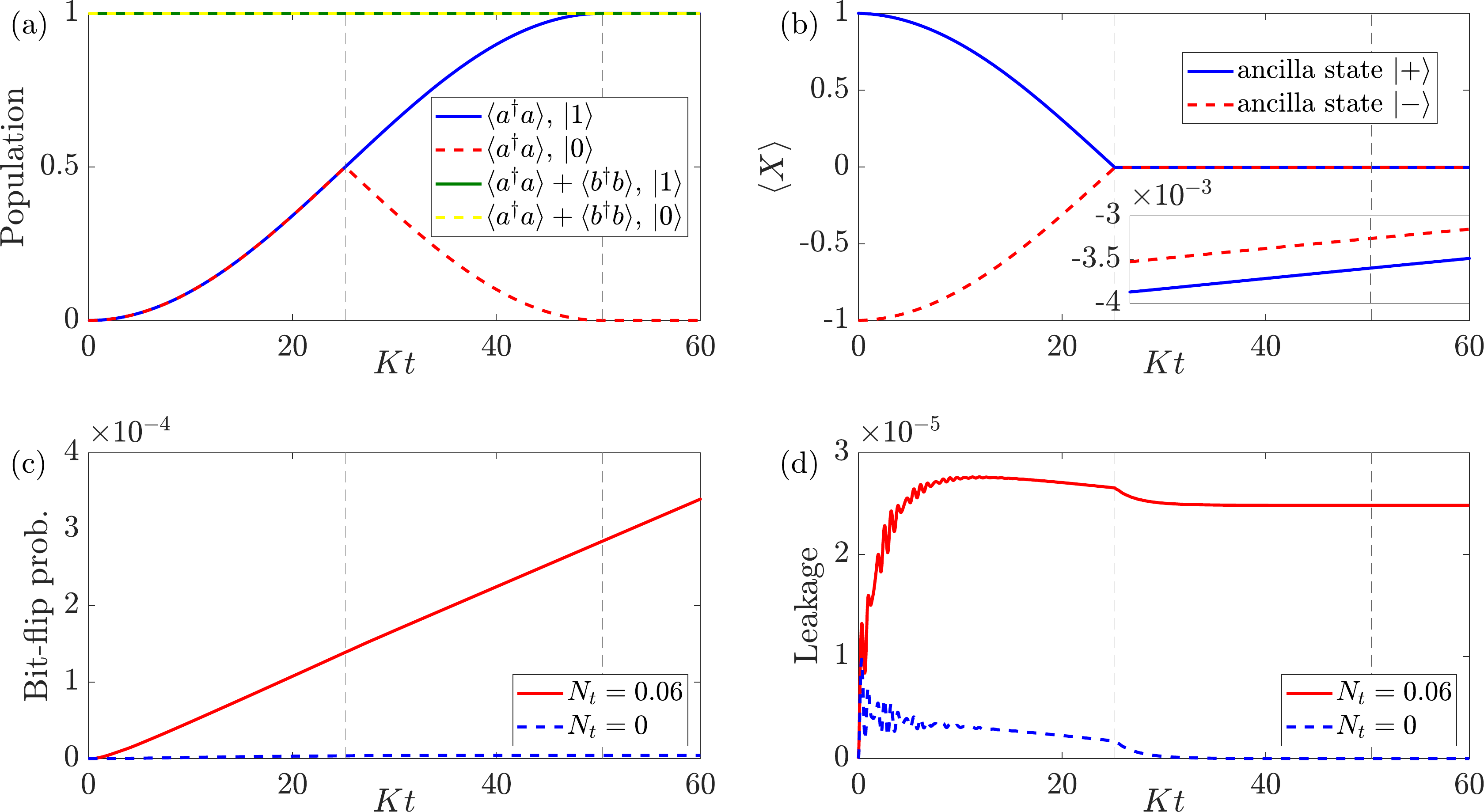}
	\caption{{Numerical simulation of the controlled-beam splitter circuit.} (a) Cavity populations, the Kerr cat (b) phase rotation, (c) bit-flip probability, and (d) leakage. The cavity fields start from the Fock state $|01\rangle$ and the various initial states of the Kerr-cat qubit are indicated in the plot legends. In the simulations, the cPBS and BS interactions are applied sequentially starting with the cPBS coupling. The time when we switch from the cPBS coupling to the BS coupling is indicated by the first vertical dashed line. The second dashed vertical line is when the BS coupling is turned off. The experimentally realistic parameters \cite{Grimm2020} (see also Table~\ref{table}) we used for the simulation are $\alpha=\sqrt{3}$, $\chi/K = 0.09$, {$\zeta_1/K = 0.018 e^{-i\pi/2}$,} $\zeta_2 = \zeta_1\alpha$ (we chose this value arbitrarily so that both cPBS and BS gate durations are equal), $N = 1$ (to compensate the mean-field Stark shift), $N_t = 0.06$, {$\kappa/K = 2.0\times 10^{-4}$}, and $\kappa_2/K = 8.0\times 10^{-2}$.  
	The blue curves in panels (c,d) show the bit-flip and leakage probabilities in the absence of thermal population ($N_t = 0$), illustrating that these errors are dominated by thermal noise.
	\label{fig:TimeEvolution}}
\end{center}
\end{figure}

We have simulated the time evolution of the full system during the cBS gate (see Appendix~\ref{app:Num} for details) and plotted the results in Fig.~\ref{fig:TimeEvolution}. The two cavity fields start in the Fock state $|01\rangle$ and the Kerr-cat qubit in one of the logical qubit states {$\ket{0},\ket{1},\ket{\pm}$}. 
First, the cavity population [panel (a)] clearly shows that the combination of cPBS and BS interactions leads to the desired {cBS}  interaction: the population of the mode A increases from zero to half a photon during the cPBS, regardless of the state of the ancilla. The subsequent BS interaction then brings the population either to unity (for the ancilla in the state $|1\rangle$) or back to zero (for ancilla state $|0\rangle$), clearly showing the different phase acquired by the cavity fields during the cPBS interaction for the two ancilla states. Since the total population of the two cavities remains constant during the interaction, the population of the cavity mode B moves in the opposite direction.

The partial swap of the cavity fields during the cPBS interaction imparts back action on the ancilla which is shown in panel (b). Swapping photons between the two cavity fields leads to $Z$ 
rotation of the ancilla by an amount that depends on the number of photons and direction of swapping. Starting in one of the eigenstates of the Pauli $X$ operator $|\pm\rangle$, the cat qubit becomes entangled with the cavity fields after the cPBS interaction, bringing the expectation of the Pauli $X$ operator to zero.
In addition to this deterministic rotation, the system suffers from phase errors caused by photon loss and gain. Limited validity of the mean-field approximation (which we used to explain the dynamics but not in numerical simulations) results in a slight under- or over-rotation of the state which can be seen in the inset of Fig.~\ref{fig:TimeEvolution}(b) where the phase rotation after the cPBS is not exactly 0.

Finally, photon gain processes in the Kerr cat (associated with the action of the creation operator $\hat{c}^\dagger$) lead to bit flips and leakage which are plotted in Fig.~\ref{fig:TimeEvolution}(c,d).
The bit flip probability increases steadily over time and is largely unaffected by the gate operation.
Leakage out of the qubit subspace, on the other hand, has nontrivial dynamics during the cPBS interaction and then reaches a steady state set by the competition between photon gain $\mathcal{D}[\hat{c}^\dagger]\hat{\rho}$ and two-photon dissipation $\mathcal{D}[\hat{c}^2]\hat{\rho}$.
The fast oscillations in the leakage are caused by the sudden switching of the cPBS interaction and can be reduced by shaping the pumps to a more adiabatic profile. The overall higher leakage rate compared to the steady state is caused by the cPBS Hamiltonian itself, specifically the term $\hat{a}^\dagger\hat{b}\hat{c}^\dagger$ which {can take} the Kerr cat out of the qubit subspace whenever a photon is swapped from mode B to mode A.
{Since the bit-flip and leakage errors are the same for both logical states of the ancilla, we expect them to be the same for all possible ancilla states as well.}

To get a complete picture of the gate performance, we characterize it with quantum process tomography. We use Fock encoding  in the cavities to define qubits.  {That is, the logical states of the cavity are simply the two lowest boson number states: $|0_\mathrm{L}\rangle=|n=0\rangle, |1_\mathrm{L}\rangle=|n=1\rangle$.} We simulate the evolution of all three-qubit Pauli operators of the whole system, $\vec{P} = (I_1I_2I_3, I_1I_2X_3, I_1I_2Y_3, \ldots, Z_1Z_2Z_3)^T$, where $I_j$ is the identity and the operators act, in turn, on the cavity mode $a$, mode $b$, and the Kerr cat. With these results, we can then formulate the Pauli transfer matrix $R$ of the three-qubit system. Leakage out of the qubit subspace (corresponding to excitations of the Kerr cat out of the ground state manifold and bunching of photons in the cavities) is quantified as the deviation of the $R_{1,1}$ element of the Pauli transfer matrix from the ideal value of one  $p_{\rm leak} = 1-R_{1,1}$.

The Pauli transfer matrix we obtain can be expressed as a product of the ideal Pauli transfer matrix [obtained from the evolution governed by the qubit-subspace Hamiltonian $H_{\rm id} = \frac{1}{4}\chi\alpha(I_3-Z_3)(X_1X_2+Y_1Y_2)$ without dissipation] and a noise transfer matrix, $R = R_{\rm noise}R_{\rm id}$. From the noise transfer matrix, we then evaluate the gate fidelity 
as $F = (\mathrm{Tr}[R_{\rm noise}] +d)/(d^2 +d)$, where $d=2^n$ and $n=3$ is the number of qubits \cite{Chow2012},
and the noise process matrix $\chi_{\rm noise}$ \cite{Greenbaum2015}. The elements of the noise process matrix describe the dephasing and non-dephasing errors of the gate. The total dephasing error is calculated by adding up all the diagonal elements of the process matrix that have only Pauli $Z$ or $I$ components (not including $I_1I_2I_3$). The non dephasing error is calculated by adding up the rest of the diagonal terms (i.e., elements containing  at least one $X$ or $Y$ Pauli operator). 

\begin{figure}
\begin{center}
	\includegraphics[width=1.0\linewidth]{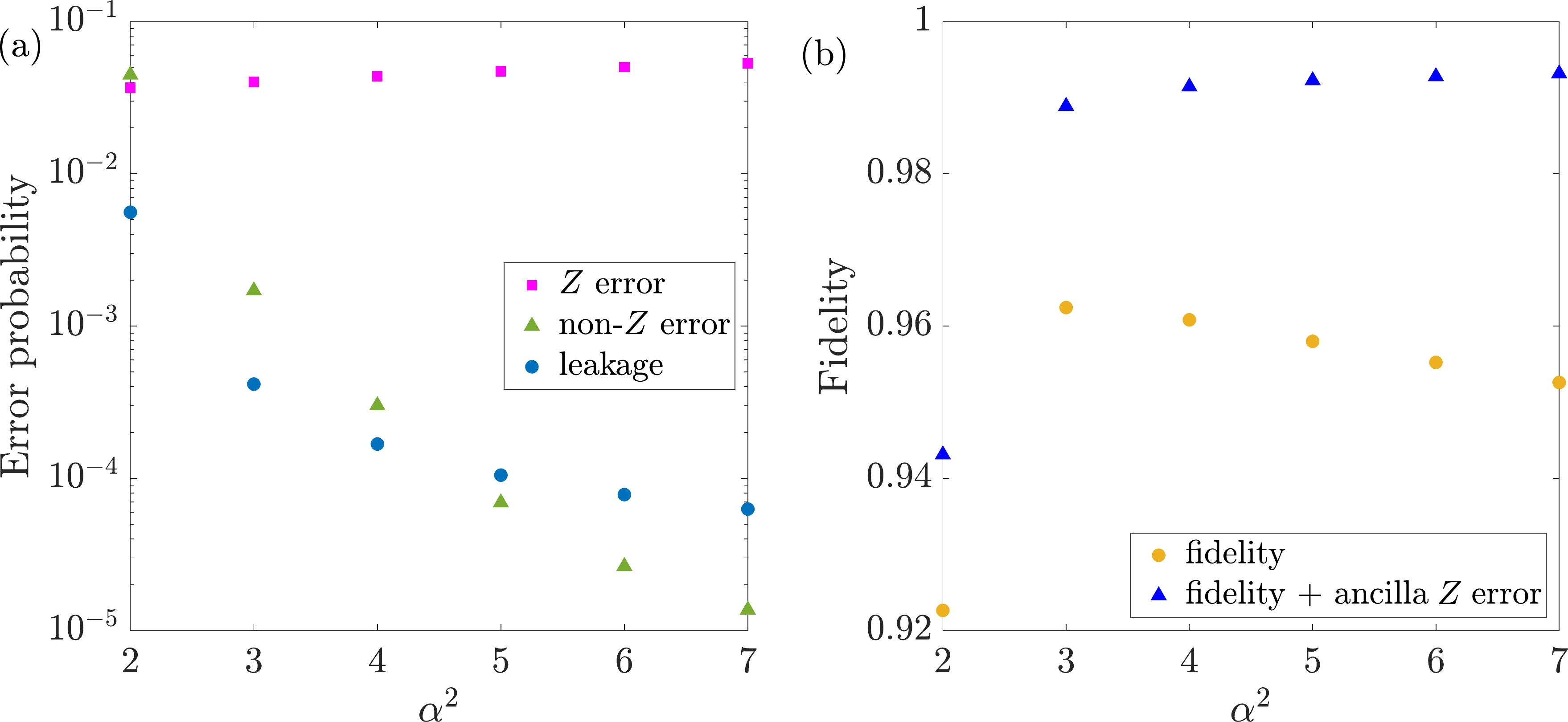}
	\caption{(a) Qubit errors (dephasing $Z$, non-dephasing non-$Z$, and leakage)  and (b) gate fidelity as a function of the cat size $\alpha^2$. Drop in fidelity for $\alpha^2=2$ is associated with non-$Z$ errors induced by the cross-Kerr coupling term in Eq.~\eqref{eq:HamEff}. {The blue triangles in (b) show the gate fidelity based on the modified noise transfer matrix in which the ancilla $Z$ errors are factored out.} 
	System parameters are the same as in Fig.~\ref{fig:TimeEvolution}.}\label{fig:Noise}
\end{center}
\end{figure}

We plot the three types of error---dephasing ($Z$), non-dephasing (non-$Z$), and leakage---against the cat size in Fig.~\ref{fig:Noise}(a).
As the cat size $\alpha$ increases, the energy gap between the cat states { $|\mathcal{C}_\pm\rangle$} and the rest of the cat space increases. This reduces the leakage from the Kerr-cat qubit subspace; at the same time, the increased height of the barrier in the double-well potential of the Kerr-cat Hamiltonian also suppresses tunnelling between the two logical states, leading to exponential reduction of bit-flip errors as well. In addition, the ancilla dephasing errors increase due to the growing excitation loss rate $\kappa\alpha^2$, becoming the dominant source of error for moderately sized cats. This increase is, however, sublinear in $\alpha^2$ as larger cat size also leads to a faster cBS gate. 
For large cat sizes, the fidelity decreases owing to the stronger ancilla dephasing but remains above 95 \% for realistic cat sizes, $\alpha^2\leq 7$ (see Fig.~\ref{fig:Noise}(b)).

{For small cat sizes, the leakage and non-dephasing errors are further enhanced by the cross-Kerr interaction between the cavity fields and the Kerr-cat qubit. This large contribution can be attributed to the different photon numbers of the logical states $\ket{0,1}$ which are not exactly equal to the coherent states $\ket{\pm\alpha}$. The cross-Kerr coupling then gives rise to a transition element $\langle 1|(c^\dagger c-\alpha^2)\ket{0} = -\alpha^2\csch(2\alpha^2)$ which, for the cavities initially in the state $\ket{00}$ or $\ket{11}$ (for which the mean-field compensation does not fully cancel the cross-Kerr coupling) gives rise to bit flip probability (from time-dependent perturbation theory) $\chi^2\alpha^4\csch^2(2\alpha^2)t^2$. For $\alpha^2 = 2$, this gives a bit-flip probability (for the two cavity input states) of about 10 \% which is consistent with the overall non-$Z$ error probability (averaged over all possible input states) of about 4.5 \% (without cross-Kerr interaction, the non-$Z$ error probability is two orders of magnitude smaller). This error quickly drops with the cat size---for $\alpha^2=7$, the bit-flip probability due to the cross-Kerr interaction is about $2.7\times 10^{-9}$, which is negligible in the total non-$Z$ error probability of $1.3\times 10^{-5}$.}

As discussed in the introduction, an advantage of the Kerr cat is that these ancilla phase-flip errors do not propagate back into the cavities.
{Because of that, it is useful to look at a measure of the fidelity that factors out the ancilla $Z$ errors. This is done with a modified noise transfer matrix, $R = \bar{R}_{\rm noise}R_{IIZ}R_{\rm id}$, where we separate the ancilla $Z$ errors from the rest of the noise. Here $R_{IIZ} = I\otimes I\otimes{\rm diag}\{1,1-2p,1-2p,1\}$, where $p = \kappa\alpha^2 t$ is the ancilla phase flip probability and $t$ is the gate time. The fidelity is then evaluated using the noise transfer matrix $\bar{R}_{\rm noise}$. The modified fidelity  increases with the cat size and for larger cat sizes is above 99 \% (see Fig.~\ref{fig:Noise}(b)). For $\alpha^2 = 7$ the value is 99.3 \% compared to the fidelity (including $Z$ errors) of 95.3 \% which shows that the largest contribution to gate infidelity is indeed from ancilla $Z$ errors for which the gate is transparent.}

Experimental process tomography of the three-qubit gate would be extremely ineffective, especially if one were to estimate the rare bit-flip and leakage errors. An easier way to estimate these errors is by evaluating photon bunching based on the following argument: Phase-flip errors only change the overall phase of a state but not the splitting ratio which is set by the length of the cPBS and BS gates. Bit flips, on the other hand, change the splitting ratio of the cPBS gate---in the extreme case where the bit flip occurs exactly in the middle of the cPBS gate, both halves cancel each other and only the deterministic beam splitter is applied. When starting with one photon in each cavity and performing a full swap by the cBS gate, such an error would therefore lead to the two photons bunching with certainty in one of the cavity modes which can be measured with a photon-number or parity measurement.

\begin{figure}
\begin{center}
	\includegraphics[width=1.0\linewidth]{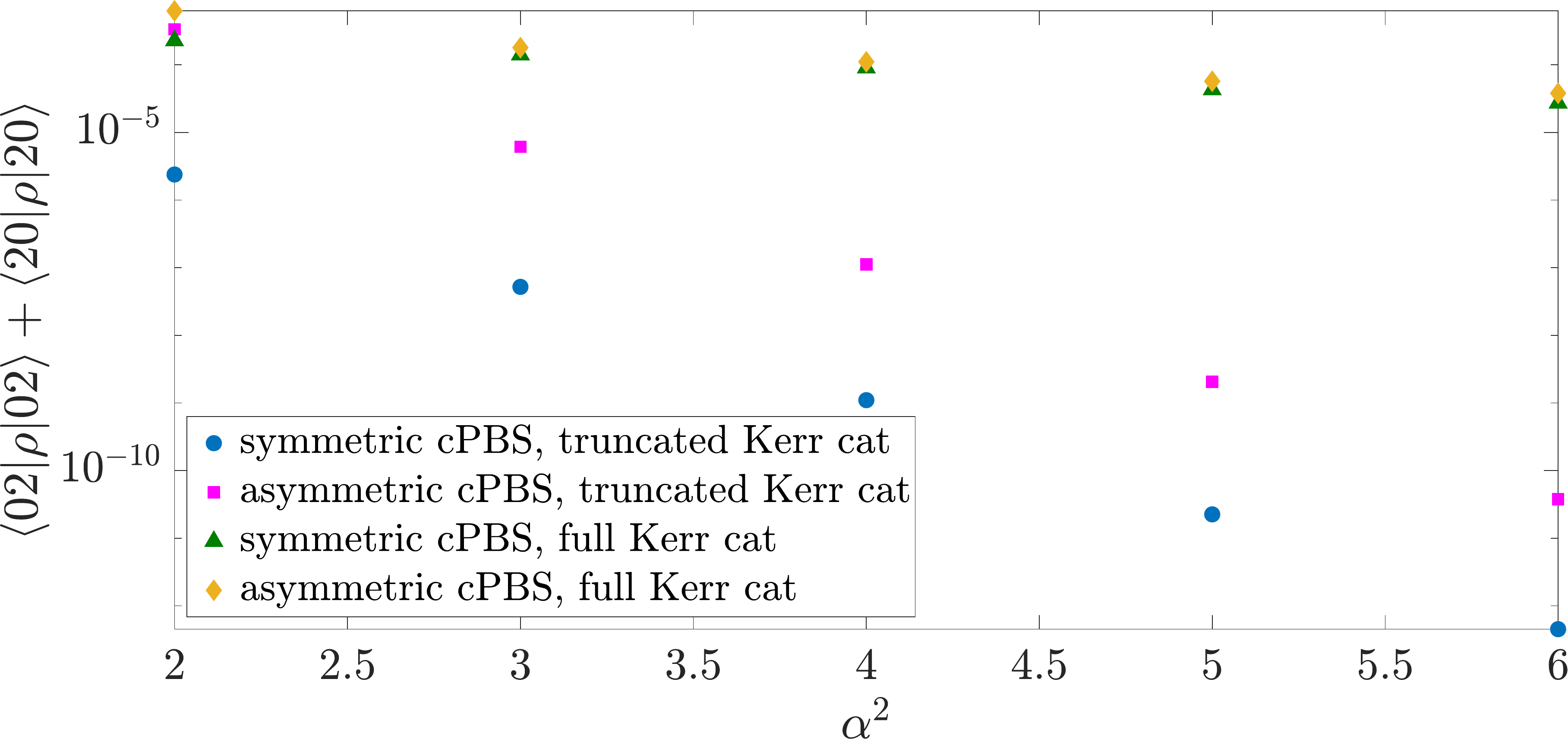}
	\caption{Cavity photon bunching as a function of the cat size $\alpha^2$. For the asymmetric cPBS we have used the Hamiltonian in Eq.~(\ref{eq:cPBS}) and Eq.~(\ref{eq:SymcPBS}) for the symmetric cPBS. In the truncated Kerr cat simulations, we have included only the two lowest eigenstates of the Kerr cat {(corresponding to the qubit logical subpace)}, removing leakage errors; the full Kerr cat simulations include also the higher states. Parameters are same as in Fig.~\ref{fig:TimeEvolution} except, for symmetric cPBS {$\zeta_1/K = 0.009 e^{-i\pi/2}$}, $\zeta_2 = 2\zeta_1\alpha$ in order to keep the gate time the same as for asymmetric cPBS.}\label{fig:Bunching}
\end{center}
\end{figure}

We simulate this effect by initializing the cavity modes in the state $|11\rangle$ and evaluating the {unwanted} population of the two-photon states in each of the cavity modes, $\ket{20}$ and $\ket{02}$ as shown in Fig.~\ref{fig:Bunching}. 
When considering only bit-flip errors and disregarding leakage (which can be achieved by using only the two qubit levels of the Kerr cat in the simulations) the probability of populating the two-photon states reduces exponentially with the cat size in agreement with the reduced probability of bit flips. With leakage included, however, the two-photon population is much larger and decreases much more slowly with $\alpha^2$. In a deep double-well potential of the Kerr cat, leakage errors can increase bunching in two ways: by modifying the beam-splitter rate (causing under- or over-rotation of the swap) or by increased tunnelling owing to the reduced potential barrier in the excited state compared to the ground state. Nevertheless, for the moderate sizes we consider here, this simple picture breaks down as only the ground state manifold is located within the double well and the excited states lie above it.


The {undesired} photon bunching can be further reduced by symmetrizing the cPBS Hamiltonian as the Hamiltonian in Eq. (\ref{eq:HamEff}) is asymmetric in the cPBS interaction. 
The operators $\hat{c}$ and $\hat{c}^\dagger$ are not exactly equivalent to the Pauli $Z$ operators in the Kerr cat qubit basis but each include a small Pauli $Y$ component as well {\cite{Puri2020}}. In addition, Eq.~\eqref{eq:HamEff} introduces leakage when swapping photons from cavity $b$ to cavity $a$ but not in the opposite direction. Changing the cPBS Hamiltonian to 
\begin{equation}\label{eq:SymcPBS}
\hat{H}_{\rm cPBS} = -(\hat{a}^\dagger\hat{b} +\hat{a}\hat{b}^\dagger)( \zeta_1\hat{c}^\dagger +\zeta_1^\ast\hat{c}) 
\end{equation}
removes both problems. First, since within the logical subspace $\hat{c} +\hat{c}^\dagger = 2\alpha Z$ is diagonal {\cite{Puri2020}}, we remove systematic bit-flip errors, reducing the corresponding bunching (cf.\ magenta squares and blue circles in Fig.~\ref{fig:Bunching}). Second, it symmetrizes leakage errors, leading to equal populations of the states $\ket{20}$ and $\ket{02}$. This symmetrization can be achieved with an additional drive applied to the device at frequency $\omega_c-\Delta$ which resonantly enhances the terms $\hat{a}^\dagger\hat{b}\hat{c} + \rm{H.c.}$ in four-wave mixing. Although such an addition is, in principle, possible, it brings the risk of increased absorption heating of the chip, leading to more errors. Moreover, the benefit it provides seems relatively minor when leakage is considered (cf. yellow diamonds and green triangles in Fig.~\ref{fig:Bunching}). It is therefore more practical to aim for the asymmetric cPBS Hamiltonian~\eqref{eq:HamEff} in near-future experiments.

\section{Conclusions}

We have presented a new interferometric method for creation of a circuit-QED beam-splitter operation between two bosonic modes that is controlled by the state of a Kerr-cat ancilla qubit.  Application of an appropriate drive tone to the Kerr-cat induces a beam-splitter Hamiltonian whose phase depends on the internal state of the Kerr-cat.  Combining this conditional phase beam-splitter with an unconditional beam-splitter yields either identity or SWAP (up to a phase in one mode) of the two bosonic modes depending on the internal state of the Kerr-cat.

A positive feature of our method is that the strong noise bias of the Kerr-cat makes the operation error transparent with respect to the dominant ancilla faults and thereby  permits a (nearly)  nondestructive measurement of the SWAP operator.  This in turn considerably simplifies protocols for stabilization of quantum computations~\cite{berthiaume1994stabilisation,barenco1997stabilization,peres1999error}, state purification~\cite{cirac1999optimal} and cooling~\cite{cotler2019quantum}. 


Naively, one might expect that the low anharmonicity in the bare SNAIL \cite{SivakSNAIL} (see Table \ref{table}) used to realize the Kerr-cat ancilla \cite{Grimm2020} could make the four-wave mixing smaller and the gate relatively slow in comparison with existing cSWAP protocols with transmons \cite{GaoYY2018,Gao2019}.  However, the anharmonicity of the Kerr-cat (the gap to states outside the code space) increases with the strength of the pump driving the SNAIL and can be considerably larger than the bare/un-pumped SNAIL's anharmonicity.  Current experiments have already demonstrated single-qubit Kerr-cat operations \cite{Grimm2020} that are as fast as transmon gates, even though the bare anharmonicity of the SNAIL used to realize the Kerr-cat is 5-10 times smaller than that of a typical transmon anharmonicity.  Additionally, the gate speed is directly proportional to the size of the cat since the cat amplitude provides one of the four waves being mixed and this `internal pump' is substantially larger than can be typically achieved with external pumping. Thus, a large amplitude Kerr-cat can, in principle, be faster than existing transmon-based cSWAP protocols.  In the transmon case, a full swap has been performed in $t \sim 10\;{\rm \mu s}$ \cite{Gao2019}. In comparison, we predict for the SNAIL setup described here that the gate time (for the sequential gate), using the Kerr nonlinearity $K/(2\pi) = 6.7\;{\rm MHz}$ from Ref.~\cite{Grimm2020}, is $t = 1.2\;{\rm \mu s}$.
In our simulations we have set the cPBS and deterministic BS gate times equal for simplicity. By setting the gate times independently, we could increase the BS drive amplitude and decrease the gate time slightly.
Another possibility for a shorter gate time is the simultaneous driving schemes discussed in Appendix \ref{app:schemes}. With these, the gate time could be halved (if certain complications discussed in Appendix \ref{app:schemes} are addressed).

Several complex experimental factors can limit the maximum pump amplitude allowed before our theoretical analysis breaks down. Predicting this parameter limit requires complex system modelling \cite{Zhang2019} that is beyond the scope of the present work. Finally, we note that while the first generation of Kerr-cat experiments has demonstrated the predicted ancilla bit-flip lifetime enhancement \cite{Grimm2020}, a new generation of experiments with still larger cats has found even greater lifetime enhancement \cite{FrattiniThesis}.  {Based on the parameter values we have assumed, the fidelity of our cSWAP gate is primarily limited by dephasing errors of the Kerr-cat ancilla associated with single excitation loss/gain and by the residual cross-Kerr interaction between the bosonic modes and the Kerr-cat.  These facts suggestion directions for future improvements in the design. }

Based on recent progress in realizing Kerr-cat qubits \cite{Grimm2020}, our proposal is experimentally feasible, requiring only a single additional drive tone (alternative driving schemes are discussed in Appendix C).  No additional non-linear elements are needed as the Kerr-cat qubit itself supplies the non-linear element and state-dependent `internal pump tone.'  Kerr-cat ancillas have already been used to create state-dependent displacements of a single bosonic mode \cite{Grimm2020}.  The next step needed to create the circuit described here is a single Kerr-cat coupled to two bosonic modes.
With these straightforward extensions of existing experimental devices, our proposal offers great potential for implementing error transparent swap operations with a broad range of applications.


\section{Acknowledgments}
{SMG and SP acknowledge support by the Air Force Office of Scientific Research under award number FA9550-21-1-0209.}
{IP and O\v{C} have received funding from the project LTAUSA19099 of the Czech Ministry of Education, Youth and Sports (MEYS \v{C}R). RF acknowledges project 21-13265X of the Czech Science Foundation. IP, O\v{C}, and RF have further been supported by the European Union’s 2020 research and innovation programme (CSA - Coordination and support action, H2020-WIDESPREAD-2020-5) under grant agreement No. 951737 (NONGAUSS).}


\appendix
\section{Derivation of the effective Hamiltonian}\label{app:Hamiltonian}
The derivation of the effective Hamiltonian (\ref{eq:HamEff}) follows the process from Ref.~\cite{Grimm2020}:
We start from the initial Hamiltonian
\begin{subequations}
\begin{equation}
\hat{H} = \hat{H}_{\rm fields} +\hat{H}_{\rm SNAIL} +\hat{H}_{\rm drive},
\end{equation}
\begin{eqnarray}
\hat{H}_{\rm fields} &=& \omega_{a,0} \hat{a}_0^\dagger\hat{a}_0 +\omega_{b,0} \hat{b}_0^\dagger\hat{b}_0 +g_a(\hat{a}_0^\dagger\hat{c}_0 +\hat{c}_0^\dagger\hat{a}_0) \nonumber\\ &&+g_b(\hat{b}_0^\dagger\hat{c}_0 +\hat{c}_0^\dagger\hat{b}_0), \\ 
\hat{H}_{\rm SNAIL} &=& \omega_{c,0} \hat{c}_0^\dagger\hat{c}_0 +g_3(\hat{c}_0^\dagger +\hat{c}_0)^3 +g_4(\hat{c}_0^\dagger +\hat{c}_0)^4, \nonumber\\ 
\\
\hat{H}_{\rm drive} &=& \sum_k (e^{-i\omega_k t}\epsilon_k\hat{c}_0^\dagger +e^{i\omega_k t}\epsilon_k^*\hat{c}_0),
\end{eqnarray}
\end{subequations}
which is identical to Eq.~\eqref{eq:Hini} except we use the subscripts 0 to remind us that it is expressed in terms of bare operators and frequencies. Through a series of transformations, we will change into a frame where the operators are dressed by the various interactions and driving fields.

First, we dress the operators by the interactions between the SNAIL and the cavity fields. We introduce the dressed operators $\hat{a}_1,\hat{b}_1,\hat{c}_1$ via $\hat{a}_0 = \hat{a}_1 -(g_a/\Delta_a)\hat{c}_1$, $\hat{b}_0 = \hat{b}_1 -(g_b/\Delta_b)\hat{c}_1$, and $\hat{c}_0 = \hat{c}_1 +(g_a/\Delta_a)\hat{a}_1 +(g_b/\Delta_b)\hat{b}_1$, where $\Delta_a =\omega_{a,0}-\omega_{c,0}$ and $\Delta_b =\omega_{b,0}-\omega_{c,0}$.
In terms of these dressed operators, the Hamiltonian can be expressed as
\begin{equation}
\begin{split}
\hat{H} &= \omega_a\hat{a}_1^\dagger\hat{a}_1 +\omega_b\hat{b}_1^\dagger\hat{b}_1 +\omega_c\hat{c}_1^\dagger\hat{c}_1 \\
&-\frac{g_a}{\Delta_a}\left(\frac{g_a^2}{\Delta_a} +\frac{g_b^2}{\Delta_b} \right)(\hat{a}_1^\dagger\hat{c}_1 +\hat{a}_1\hat{c}_1^\dagger) \\
&-\frac{g_b}{\Delta_b}\left(\frac{g_a^2}{\Delta_a} +\frac{g_b^2}{\Delta_b} \right)(\hat{b}_1^\dagger\hat{c}_1 +\hat{b}_1\hat{c}_1^\dagger) \\
&+\left( \omega_{c,0}\frac{g_a}{\Delta_a}\frac{g_b}{\Delta_b} +\frac{g_a g_b}{\Delta_a} +\frac{g_a g_b}{\Delta_b} \right)(\hat{a}_1^\dagger\hat{b}_1 +\hat{a}_1\hat{b}_1^\dagger) \\
&+g_3(\hat{f}_1^\dagger +\hat{f}_1)^3 +g_4(\hat{f}_1^\dagger +\hat{f}_1)^4 \\
&+\sum_k (e^{-i\omega_k t}\epsilon_k\hat{f}_1^\dagger +e^{i\omega_k t}\epsilon_k^*\hat{f}_1) \,,
\end{split}
\end{equation}
where we introduced the dressed frequencies $\omega_a = \omega_{a,0} +2g_a^2/\Delta_a +\omega_{c,0}g_a^2/\Delta_a^2$, $\omega_b = \omega_{b,0} +2g_b^2/\Delta_b +\omega_{c,0}g_b^2/\Delta_b^2$, $\omega_c = \omega_{c,0} -2g_a^2/\Delta_a -2g_b^2/\Delta_b +\omega_{a,0}g_a^2/\Delta_a^2 +\omega_{b,0}g_b^2/\Delta_b^2$, and the operator $\hat{f}_1 = \hat{c}_1 +(g_a/\Delta_a)\hat{a}_1 +(g_b/\Delta_b)\hat{b}_1 $.

Next, we perform the displacement transformation $\hat{a}_1 = \hat{a}_2 +\sum_k \xi_{a,k}e^{-i\omega_k t}$ and similar for $\hat{b}_1$ and $\hat{c}_1$. Defining the dressed driving fields
\begin{eqnarray}
\xi_{a,k} &=& \frac{g_a}{\Delta_a}\frac{\epsilon_k}{\omega_k-\omega_a}, \\
\xi_{b,k} &=& \frac{g_b}{\Delta_b}\frac{\epsilon_k}{\omega_k-\omega_b}, \\
\xi_{c,k} &=& \frac{\epsilon_k}{\omega_k-\omega_c},
\end{eqnarray}
we are left with the Hamiltonian
\begin{equation}
\begin{split}
\hat{H} &= \omega_a\hat{a}_2^\dagger\hat{a}_2 +\omega_b\hat{b}_2^\dagger\hat{b}_2 +\omega_c\hat{c}_2^\dagger\hat{c}_2 +g_3\hat{f}^3 +g_4\hat{f}^4 \\
&-\frac{g_a}{\Delta_a}\left(\frac{g_a^2}{\Delta_a} +\frac{g_b^2}{\Delta_b} \right)[\hat{a}_2^\dagger\hat{c}_2 +\hat{a}_2^\dagger\sum_k\xi_{c,k}e^{-i\omega_k t} \\ 
&+\hat{c}_2^\dagger\sum_k\xi_{a,k}e^{-i\omega_k t}] \\
&-\frac{g_b}{\Delta_b}\left(\frac{g_a^2}{\Delta_a} +\frac{g_b^2}{\Delta_b} \right)[\hat{b}_2^\dagger\hat{c}_2 +\hat{b}_2^\dagger\sum_k\xi_{c,k}e^{-i\omega_k t} \\ 
&+\hat{c}_2^\dagger\sum_k\xi_{b,k}e^{-i\omega_k t}] \\
& +\left( \omega_{c,0}\frac{g_a}{\Delta_a}\frac{g_b}{\Delta_b} +\frac{g_a g_b}{\Delta_a} +\frac{g_a g_b}{\Delta_b} \right)[\hat{a}_2^\dagger\hat{b}_2 \\
&+\hat{a}_2^\dagger\sum_k\xi_{b,k}e^{-i\omega_k t} +\hat{b}_2^\dagger\sum_k\xi_{a,k}e^{-i\omega_k t}] +{\rm H.c.}
\end{split}
\end{equation}
where
\begin{subequations}
\begin{align}
\hat{f} =& \hat{c}_2 +\frac{g_a}{\Delta_a}\hat{a}_2 +\frac{g_b}{\Delta_b}\hat{b}_2 \nonumber\\
&+\sum_k\xi_{k,{\rm eff}}e^{-i\omega_k t} +{\rm H.c.}, \\
\xi_{k,{\rm eff}} =& \frac{g_a}{\Delta_a}\xi_{a,k} +\frac{g_b}{\Delta_b}\xi_{b,k} +\xi_{c,k}.
\end{align}
\end{subequations}
As a final step, we move to the rotating frame with respect to the free Hamiltonian $H_0 = \omega_a'\hat{a}_2^\dagger\hat{a}_2+\omega_b'\hat{b}_2^\dagger\hat{b}_2+\omega_c'\hat{c}_2^\dagger\hat{c}_2$.
Assuming all the frequencies $\omega_a'$, $\omega_b'$, $\omega_c'$, and $\omega_k$ to be different, the only possible non-rotating terms in the Hamiltonian are
\begin{equation}\label{eq:H2}
\begin{split}
\hat{H} &= (\omega_a-\omega_a')\hat{a}^\dagger\hat{a} +(\omega_b-\omega_b')\hat{b}^\dagger\hat{b} +(\omega_c-\omega_c')\hat{c}^\dagger\hat{c} \\ 
&+g_3\hat{f}^3 +g_4\hat{f}^4 +{\rm H.c.},
\end{split}
\end{equation}
where we have dropped the subscript 2 from the creation and annihilation operators for simplicity.

Regardless of the drive frequencies, the four-wave mixing term $\hat{f}^4$ always gives the non-rotating terms
\begin{equation}\label{eq:NonRotating}
\begin{split}
&6g_4(\hat{c}^{\dagger 2}\hat{c}^2 +2\hat{c}^\dagger\hat{c})+6g_4\frac{g_a^4}{\Delta_a^4}(\hat{a}^{\dagger2}\hat{a}^2 +2\hat{a}^\dagger\hat{a}) \\
&+6g_4\frac{g_b^4}{\Delta_b^4}(\hat{b}^{\dagger2}\hat{b}^2 +2\hat{b}^\dagger\hat{b}) +12g_4\frac{g_a^2}{\Delta_a^2}(2\hat{a}^\dagger\hat{a}\hat{c}^\dagger\hat{c} +\hat{a}^\dagger\hat{a} +\hat{c}^\dagger\hat{c}) \\
&+12g_4\frac{g_b^2}{\Delta_b^2}(2\hat{b}^\dagger\hat{b}\hat{c}^\dagger\hat{c} +\hat{b}^\dagger\hat{b} +\hat{c}^\dagger\hat{c}) \\
&+12g_4\frac{g_a^2}{\Delta_a^2}\frac{g_b^2}{\Delta_b^2}(2\hat{a}^\dagger\hat{a}\hat{b}^\dagger\hat{b} +\hat{a}^\dagger\hat{a} +\hat{b}^\dagger\hat{b}) \,.
\end{split}
\end{equation}
Including only the terms up to second order in $g_i/\Delta_i$, and setting the frequencies
\begin{subequations}
\begin{align}
    \omega_a' &= \omega_a +12g_4(g_a/\Delta_a)^2(1 +2\sum_k\vert\xi_{c,k}\vert^2 ),\\
    \omega_b' &= \omega_b +12g_4(g_b/\Delta_b)^2(1 +2\sum_k\vert\xi_{c,k}\vert^2),\\
    \omega_c' &= \omega_c +12g_4[1 +(g_a/\Delta_a)^2 +(g_b/\Delta_b)^2 +2\sum_k\vert\xi_{c,k}\vert^2 ],
\end{align}
\end{subequations}
we obtain the free Hamiltonian
\begin{equation}
\hat{H}_0 = 6g_4\hat{c}^{\dagger 2}\hat{c}^2 +24g_4(\frac{g_a^2}{\Delta_a^2}\hat{a}^\dagger\hat{a}\hat{c}^\dagger\hat{c} +\frac{g_b^2}{\Delta_b^2}\hat{b}^\dagger\hat{b}\hat{c}^\dagger\hat{c})
\end{equation}
which is always present independent of the driving tones. In addition to this, the resonances between the drives and the frequencies of the three modes can create additional three- and four-wave mixing terms via the last two terms in Eq.~\eqref{eq:H2}.

To implement the controlled beam splitter gate with a Kerr-cat ancilla, we need three driving tones:
The first one at the frequency $\omega_1 = 2\omega_c'$ introduces two-photon driving of the SNAIL via three-wave mixing,
\begin{equation}
\hat{H}_1 = 3g_3(\xi_{1,\rm eff}\hat{c}^{\dagger 2} +\xi_{1,\rm eff}^* \hat{c}^{2}).
\end{equation}
The controlled-phase beam splitter is implemented with a drive at the frequency $\omega_2 = \omega_c' +\Delta$, where $\Delta = \omega_a'-\omega_b'$, which introduces the four-wave mixing term
\begin{equation}
\hat{H}_{\rm cPBS} = -4K\frac{g_a}{\Delta_a}\frac{g_b}{\Delta_b}(\xi_{2,\rm eff} \hat{a}^\dagger\hat{b}\hat{c}^\dagger +\xi_{2,\rm eff}^*\hat{a}\hat{b}^\dagger\hat{c})\,,
\end{equation}
where $K = -6g_4$. Finally, the deterministic beam splitter is implemented with a drive at the frequency $\omega_3 = \Delta$ via three-wave mixing,
\begin{equation}\label{eq:BS}
\hat{H}_{\rm BS} = 6g_3\frac{g_a}{\Delta_a}\frac{g_b}{\Delta_b}(\xi_{3,\rm eff}\hat{a}^\dagger\hat{b} +\xi_{3,\rm eff}^*\hat{a}\hat{b}^\dagger).
\end{equation}
{The total Hamiltonian is then 
\begin{equation}
\begin{split}
\hat{H} &= \hat{H}_0 +\hat{H}_1 +\hat{H}_{\rm cPBS} +\hat{H}_{\rm BS} \\
&=-K\hat{c}^{\dagger 2}\hat{c}^2 -4K(\frac{g_a^2}{\Delta_a^2}\hat{a}^\dagger\hat{a} +\frac{g_b^2}{\Delta_b^2}\hat{b}^\dagger\hat{b})\hat{c}^\dagger\hat{c} \\
&+3g_3(\xi_{1,\rm eff}\hat{c}^{\dagger 2} +\xi_{1,\rm eff}^* \hat{c}^{2}) \\
&-4K\frac{g_a}{\Delta_a}\frac{g_b}{\Delta_b}(\xi_{2,\rm eff} \hat{a}^\dagger\hat{b}\hat{c}^\dagger +\xi_{2,\rm eff}^*\hat{a}\hat{b}^\dagger\hat{c}) \\
&+6g_3\frac{g_a}{\Delta_a}\frac{g_b}{\Delta_b}(\xi_{3,\rm eff}\hat{a}^\dagger\hat{b} +\xi_{3,\rm eff}^*\hat{a}\hat{b}^\dagger) \,.
\end{split}
\end{equation}

}
Defining $\epsilon = 3g_3\xi_{1,\rm eff}$, $\zeta_2 = 6g_3\frac{g_a}{\Delta_a}\frac{g_b}{\Delta_b}\xi_{3,\rm eff}$, and $\zeta_1 = 4K\frac{g_a}{\Delta_a}\frac{g_b}{\Delta_b}\xi_{2,\rm eff}$, the terms are the same as in Eq.~(\ref{eq:HamEff}). 

\section{Mean-field compensation of cross-Kerr interactions}\label{app:Kerr}
The interaction term
\begin{equation}
\hat{H}_{\rm cK} = -(\chi_{a} \hat{a}^\dagger\hat{a} +\chi_{b} \hat{b}^\dagger\hat{b})\hat{c}^\dagger\hat{c} \,,
\end{equation}
where $\chi_i = 4Kg_i^2/\Delta_i^2$, describes cross-Kerr interaction between the SNAIL and the cavity fields. Since the SNAIL is used to create a Kerr-cat qubit, its photon number is centered around $|\alpha|^2$, which can be subtracted from the cross-Kerr interaction by suitable frequency change.
By changing the rotating frame frequencies for the fields as $\omega_a' \to \omega_a' -\chi_a|\alpha|^2$, $\omega_b' \to \omega_b' -\chi_b|\alpha|^2$, we get the cross-Kerr term
\begin{equation}
\hat{H}_{\rm cK} = -(\chi_{a} \hat{a}^\dagger\hat{a} +\chi_{b} \hat{b}^\dagger\hat{b})(\hat{c}^\dagger\hat{c}-|\alpha|^2) \,.
\end{equation}
Similarly, we subtract the average population of the cavity fields, with the change of rotating frequency of the SNAIL $\omega_c' \to \omega_c' -\chi_a N_a -\chi_b N_b$. This results in the cross-Kerr term
\begin{equation}
\hat{H}_{\rm cK} = -[\chi_{a} (\hat{a}^\dagger\hat{a} -N_a) +\chi_{b} (\hat{b}^\dagger\hat{b} -N_b)](\hat{c}^\dagger\hat{c}-|\alpha|^2) \,,
\end{equation}
where $N_a$ and $N_b$ are the average photon populations of the fields. If $\chi_a=\chi_b \equiv \chi$, we end up with the mean-field compensated cross-Kerr interaction
\begin{equation}
\hat{H}_{\rm cK} = -\chi (\hat{a}^\dagger\hat{a} + \hat{b}^\dagger\hat{b} -N)(\hat{c}^\dagger\hat{c}-|\alpha|^2) \,,
\end{equation}
where $N = N_a +N_b$ is the total average population of the fields.

\section{Alternative driving schemes}\label{app:schemes}

There are many three- and four-wave mixing processes available to engineer the desired interactions. Often, this versatility serves as an advantage---we can, for example, choose whether to pump two-photon driving using a single pump tone at frequency $2\omega_c$ (three-wave mixing) or two pumps whose frequencies add up to $2\omega_c$ (four-wave mixing). Some of these processes, however, provide important limitations on the quality of target interactions.

\begin{figure}
\includegraphics[width=1.0\linewidth]{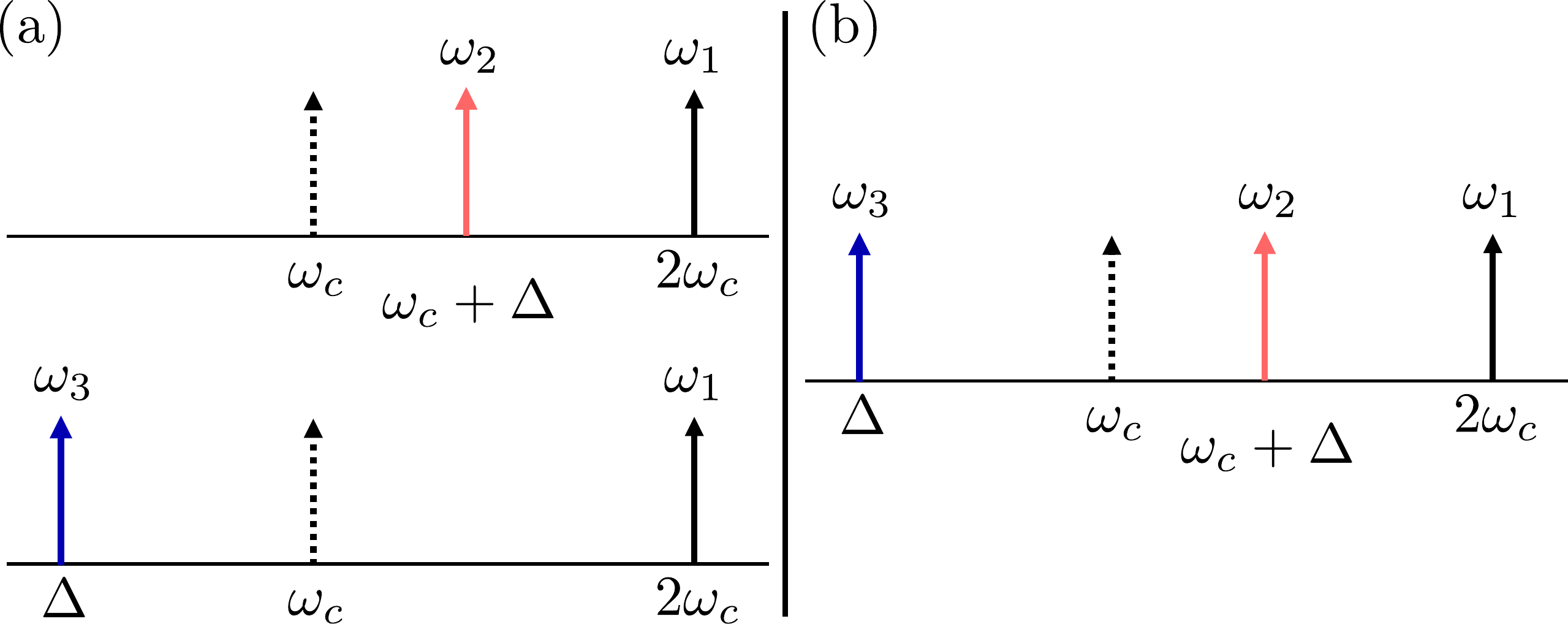}
\caption{\label{fig:drives}Different driving schemes. The drive at frequency $\omega_1 = 2\omega_c$ is used to generate the Kerr cat using 3-wave mixing as in panel (a) of Fig.~\ref{fig:mixing}. (a) Sequential driving where first the cPBS coupling is generated using 4-wave mixing with a drive at frequency $\omega_2 = \omega_c +\Delta$ as in panel (c) of Fig.~\ref{fig:mixing} where $\Delta=\omega_a-\omega_b$. After that we switch to a drive at frequency $\omega_3 = \Delta$ which creates the BS coupling as a 3-wave mixing term. (b) Simultaneous driving where both the drives at $\omega_2$ and $\omega_3$ are applied simultaneously.  
}
\end{figure}

This issue can be well illustrated {by considering if it is possible to speed up the cBS gate by simultaneously driving the} cPBS and BS interactions. In this situation (the driving frequencies are shown schematically in Fig.~\ref{fig:drives}) there are two additional resonances $\omega_2-\omega_3-\omega_c = 0$ and $\omega_1-\omega_2+\omega_3-\omega_c=0$ which introduce linear driving of the SNAIL mode. These terms do not appear in the sequential scheme discussed in the main text since they require drives at both frequencies $\omega_2$ and $\omega_3$ to be present at the same time. The total Hamiltonian then reads
\begin{equation}\label{eq:HamSim}
\begin{split}
\hat{H} &= -K\hat{c}^{\dagger 2}\hat{c}^2 +\epsilon\hat{c}^{\dagger 2} +\epsilon^* \hat{c}^{2} \\ &-\chi (\hat{a}^\dagger\hat{a} +\hat{b}^\dagger\hat{b} -N)(\hat{c}^\dagger\hat{c} -|\alpha|^2) \\
&-\frac{g_a}{\Delta_a}\frac{g_b}{\Delta_b}[\hat{a}^\dagger\hat{b}(4K\xi_{2,\rm eff}\hat{c}^\dagger -6g_3\xi_{3,\rm eff})  \\
&+\hat{a}\hat{b}^\dagger(4K\xi_{2,\rm eff}^*\hat{c} -6g_3\xi_{3,\rm eff}^*)] \\
&+6g_3(\xi_{2,\rm eff}^*\xi_{3,\rm eff}\hat{c} +\xi_{2,\rm eff}\xi_{3,\rm eff}^*\hat{c}^\dagger) \\
&-4K(\xi_{1,\rm eff}^*\xi_{2,\rm eff}\xi_{3,\rm eff}^*\hat{c} +\xi_{1,\rm eff}\xi_{2,\rm eff}^*\xi_{3,\rm eff}\hat{c}^\dagger),
\end{split}
\end{equation}
where the last two lines show the linear drive on the SNAIL.

\begin{figure}
\begin{center}
	\includegraphics[width=1.0\linewidth]{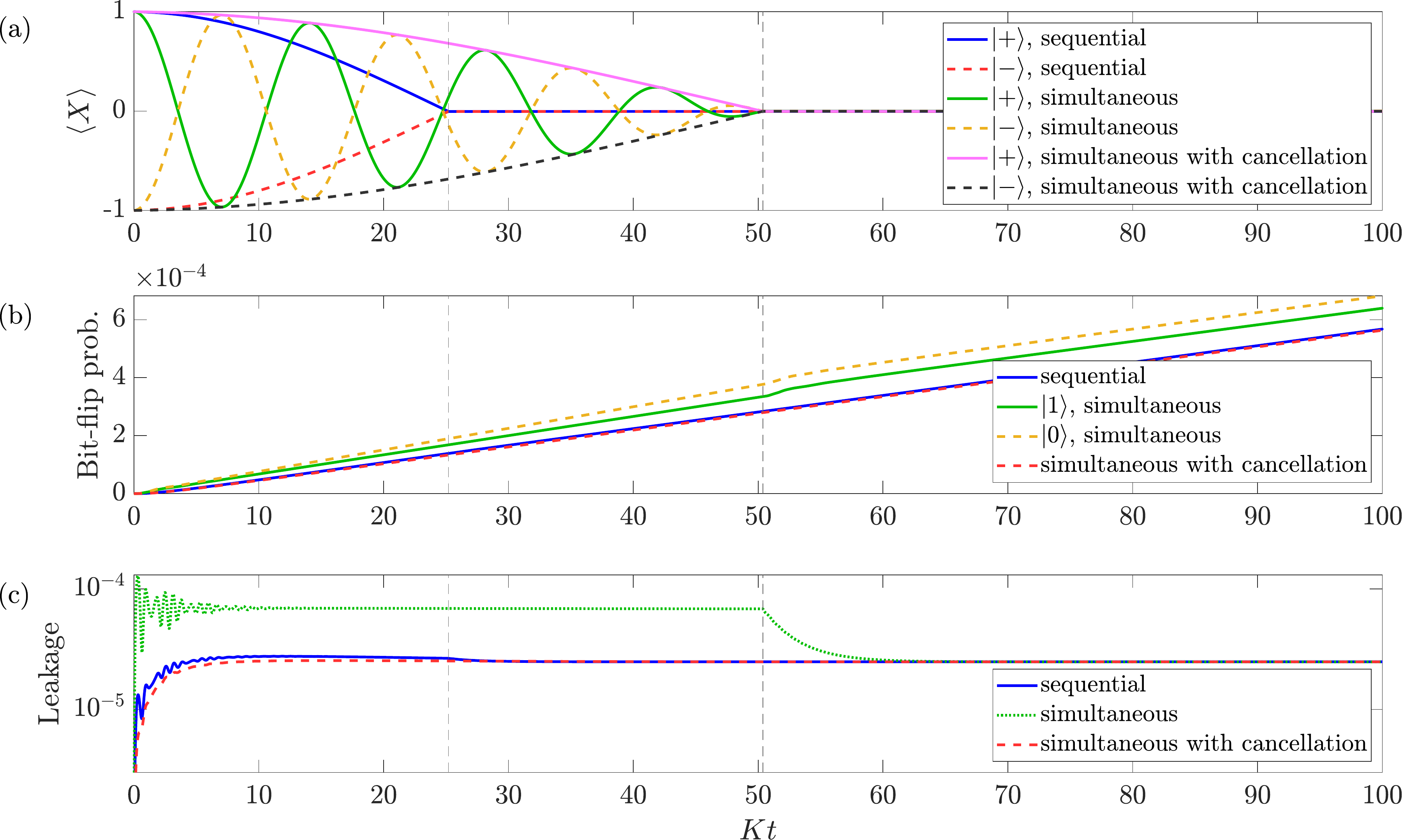}
	\caption{{ Numerical simulation of alternative driving schemes for the controlled beam splitter.} (a) Phase rotation, (b) Bit-flip probability, and (c) leakage of the Kerr-cat qubit. The curves for sequential cPBS and BS coupling are the same as in Fig.~\ref{fig:TimeEvolution}, the simultaneous lines are calculated with both the cPBS and BS drives on simultaneously using Hamiltonian (\ref{eq:HamSim}), and the simultaneous with cancellation are the same except without the linear terms in the last two lines in Hamiltonian (\ref{eq:HamSim}). For the simultaneous scheme, we have the same parameters as in Fig.~\ref{fig:TimeEvolution} except for {$\zeta_1/K = 0.009 e^{-i\pi/2}$} and $\zeta_2 = \zeta_1\alpha$ which ensures that the total gate time is the same as in the sequential case. For the linear term, we have used the value {$-0.037K\alpha (\hat{c}^\dagger +\hat{c})$}, calculated using $g_3/K=3.0$. First vertical dashed line indicates the time when the cPBS coupling is switched off and the BS coupling is switched on in the sequential case. The second vertical dashed line is when all the coupling are turned off in the sequential and simultaneous cases.} \label{fig:TimeEvolution_Drives}
\end{center}
\end{figure}

These additional linear drive terms cause $Z$ rotation of the ancilla Kerr-cat qubit and extra leakage out of the qubit subspace via the terms containing $\hat{c}^\dagger$. 
These effects can be seen in Fig.~\ref{fig:TimeEvolution_Drives} where we plot the expectation value of the Pauli $X$ operator and bit flip and leakage errors of the Kerr-cat qubit. The fast oscillations of the Pauli $X$ operator are slowly reduced as the back action from the cPBS interaction rotates the Kerr cat to a state where it is insensitive to these driving terms. This deterministic rotation can, in principle, be taken into account in designing the gate but the strength of the linear drive gives rise to significant increase in leakage and bit flip errors as well.
In addition, the bit-flip probability rises approximately linearly in time, indicating a constant rate of errors, whereas the leakage probability is quasi-stationary, reflecting the competition between the leakage rate and two-photon cooling. The small dependence of the bit-flip rate on the ancilla state for the simultaneous protocol seen in the upper two curves in panel (b) is a result of the uncancelled linear drive on the ancilla (last two lines in Eq.~\eqref{eq:HamSim}). For the other two protocols, the bit-flip probability is independent of the ancilla state.

The errors introduced by this linear term in the Hamiltonian can be compensated by an additional drive at the SNAIL resonance with the appropriate amplitude and phase. Such a pump tone would, however, contribute to absorption heating of the SNAIL device and to possible multiphoton transitions which are not captured by our simple effective model. These issues could be partially avoided by using two SNAIL devices---one, serving as the cat-qubit ancilla, would be used to engineer the cPBS interaction while the other would only provide the BS coupling. In this setting, interference between the different drives would be avoided at the cost of increasing the experimental complexity since both SNAIL devices would need to be coupled to both cavity fields and independently calibrated and controlled.

Removing the linear term from the Hamiltonian (within our effective model, by the additional linear drive on the SNAIL or by using two SNAIL devices) results in errors comparable to the sequential scheme discussed in the main text as is also shown in Fig.~\ref{fig:TimeEvolution_Drives}. Cancellation of the linear terms would then enable us to shorten the gate time by increasing the drive strengths. Detailed analysis of the resulting error budget would then require us to go beyond the effective model used in this manuscript to analyze in detail the effects of multiphoton transitions and counterrotating terms in the full initial Hamiltonian.



Finally, we note that there are other possible driving schemes that would result in the required cPBS and BS couplings using, for example, four-wave mixing for the BS interaction. However, all these alternative approaches introduce similar linear terms in the Kerr-cat Hamiltonian when the cPBS and BS interactions are switched on simultaneously. The above discussion thus applies to all of them.



\section{Numerical simulations}\label{app:Num}

In numerical simulations, we work with Kerr cats with real amplitude, $\alpha\in\mathbb{R}$. Since we include two-photon damping to help stabilize the Kerr cat in the qubit subspace, we therefore need to slightly modify the amplitude and phase of the two-photon drive.
The {no-jump part of the} two-photon dissipation can be thought of as an additional non-Hermitian term {in the Kerr-cat Hamiltonian}, $-i\kappa_2\hat{c}^{\dagger 2}\hat{c}^2/2$, changing the cat amplitude to $\alpha = \sqrt{\epsilon/(K +i\kappa_2/2)}$. In order to keep the cat amplitude real, we adjust the two-photon drive $\xi_1$ at frequency $\omega_1$ such that $\epsilon = \alpha^2 \sqrt{K^2+\kappa_2^2/4}\, e^{i\phi}$, where $\phi = \tan^{-1}[\kappa_2/(4K)]$.


\begin{figure}
\begin{center}
	\includegraphics[width=1.0\linewidth]{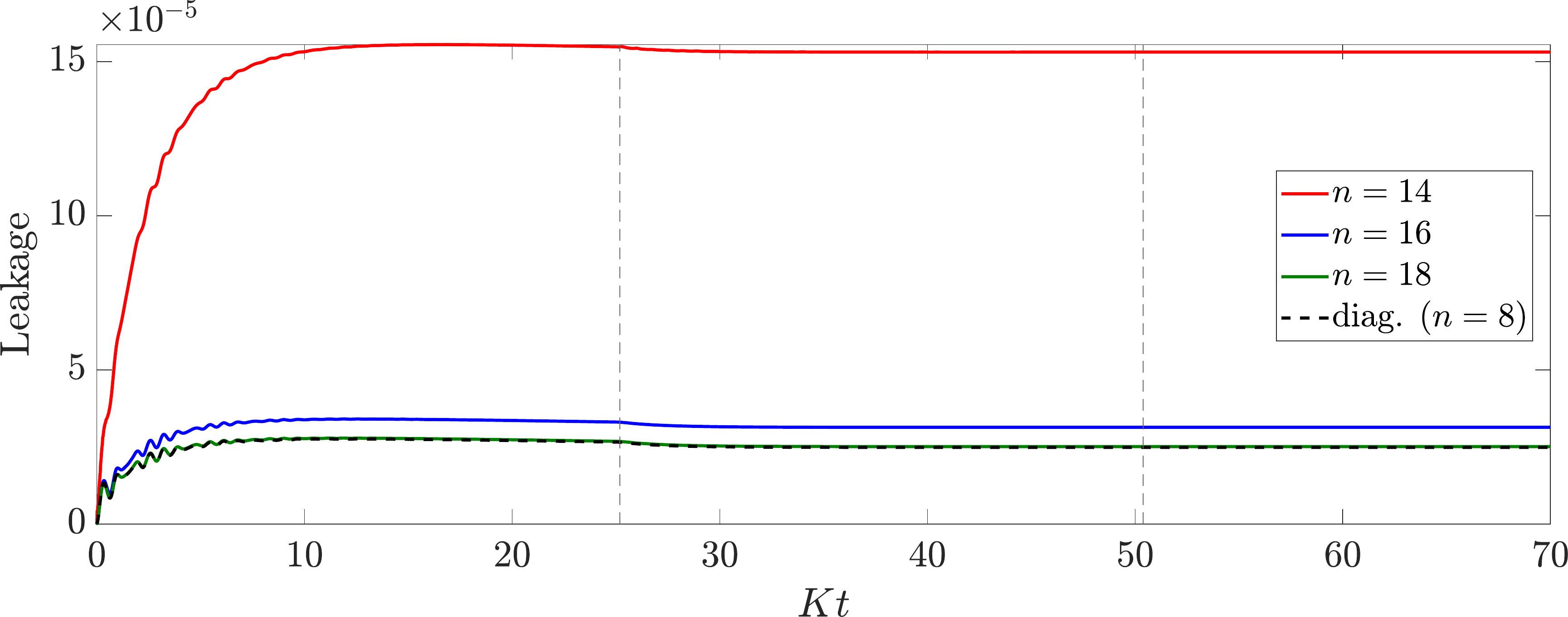}
	\caption{Leakage {convergence} of the Kerr cat {simulations} computed with different truncations of the Kerr-cat Hilbert space sizes of $n=14, 16, 18$ (in the Fock basis). The black dashed line is obtained with the Kerr cat in the diagonal basis where only 8 states in the cat Hilbert space are needed. The cavities are initially in the Fock state $|01\rangle$. The time when we switch from the cPBS coupling to the BS coupling is indicated by the first vertical dashed line. The second dashed vertical line is when the BS coupling is turned off. The parameters are the same as in Fig.~\ref{fig:TimeEvolution}.}\label{fig:DiagCat}
\end{center}
\end{figure}

We perform numerical simulations of the system in Python using the software package QuTiP \cite{Johansson2012}.
The calculations are done in the eigenbasis of the Kerr cat instead of its Fock basis to make the simulations faster. The low-lying states in the inverted double-well potential of the Kerr cat are approximately given by the displaced Fock states $\ket{n(\pm\alpha)}=D(\pm\alpha)\ket{n}$, where $D(\alpha) = \exp(\alpha\hat{c}^\dagger-\alpha^\ast \hat{c})$ is the displacement operator~\cite{Puri2019}. In simulations, we numerically diagonalize the Hamiltonian $\hat{H}_{\rm diag} = -K\hat{c}^{\dagger 2}\hat{c}^2 +\varepsilon\hat{c}^{\dagger 2} +\varepsilon^* \hat{c}^{2}$ and transform the operators $\hat{c}$ and $\hat{c}^\dagger$ in this new basis. To ensure that the two-photon dissipation is correctly taken into account, the amplitude of the two-photon drive in $\hat{H}_{\rm diag}$ is set such that $\varepsilon/K = \epsilon/(K+i\kappa_2/2)$.

This approach allows us to work with a much smaller Hilbert space for the Kerr-cat ancilla. This effect can be seen in Fig.~\ref{fig:DiagCat} where we plot the leakage error of the ancilla for different Hilbert space sizes with the Fock encoding (solid lines). We need to include at least 18 Fock states in the ancilla to obtain the correct leakage error; in the diagonal basis, only the first 8 states are sufficient to get the same result (dashed black line). {This difference between the Fock and diagonal bases is even more striking for larger cat sizes: we estimate that about 40 Fock states would be needed to simulate a Kerr cat with $\alpha^2 = 7$ photons while only the first 12 states in the diagonal basis are sufficient.}

\end{document}